\pdfoutput=1
\documentclass[12pt]{article} 
\usepackage{sastyle} 
\usepackage{marvosym}
\graphicspath{ {images/} }

\newcommand*\Ampskip{8mu}
\catcode`,\active
\newcommand*\Amp{\begingroup
	\catcode`\,\active
	\def ,{\mskip\Ampskip\relax}%
	\doAmp
}
\catcode`\,12
\def\doAmp#1#2#3#4{%
	\mathcal{A}^{#1}_{#2}\biggl[\genfrac..{0pt}{}{#3}{#4}\biggr]%
	\endgroup
}
\begin{document}
	\vspace*{-.6in} \thispagestyle{empty}
	\begin{flushright}
	\end{flushright}
	\vspace{.2in} {\Large
		\begin{center}
			{\bf On loop celestial amplitudes	for gauge theory and gravity
				\vspace{.1in}}
		\end{center}
	}
	\vspace{.2in}
		\begin{center}
			{\bf 
				Soner Albayrak$^\text{\Mars}$, Chandramouli Chowdhury$^\text{\Moon}$, and Savan Kharel$^\text{\Sun,\Neptune}$}
			\\
			\vspace{.2in} 
		$^\text{\Mars}$ {\it  Department of Physics, Yale University, New Haven, CT 06511}\\
		$^\text{\Moon}$  {\it ICTS, Tata Institute of Fundamental Research, Sivakote, Bangalore 560089, INDIA}\\
		$^\text{\Sun}$ {\it  Department of Physics, Williams College, Williamstown, MA 01267} \\
		$^\text{\Neptune}$ {\it  Department of Physics,  University of Chicago, Chicago, IL 60637}
		\end{center}
		
		\vspace{.2in}
		
	\begin{abstract}
Scattering amplitudes of massless particles in Minkowski space can be expressed in a conformal basis by Mellin transforming the momentum space amplitudes to correlation functions on the celestial sphere at null infinity.
In this paper, we study celestial amplitudes of loop level gluons and gravitons. We focus on the rational amplitudes that carry all-plus and single-minus external helicities. Because these amplitudes are finite, they provide a concrete example of celestial amplitudes of Yang- Mills and gravity theory beyond tree level. We give explicit examples of four and five point functions and comment on higher point amplitudes.
	\end{abstract}
	
	\newpage
	
	\tableofcontents
	
%	\newpage
	
% BODY
\section{Introduction}
The scattering amplitudes in Minkowski space can be mapped to the celestial sphere at light-like infinity, where they are encoded in terms of conformal correlators \cite{Pasterski:2016qvg}.  These correlation functions go by the ethereal name of \emph{celestial amplitudes} and they exhibit conformal symmetry at the boundary for bulk observables. This observation provides a complementary representation of scattering amplitudes where they are beheld as a holographically dual conformal field theory residing in the celestial sphere. Thus, the holographic nature of celestial amplitudes in principle can shine light to an outstanding problem, i.e. what is a concrete holographic formulation for flat spacetime?\footnote{ Flat space holography was already proposed in \cite{deBoer:2003vf} where Minkowski space was foliated along the Euclidean Anti-de Sitter and de Sitter slices. Other related approach, please refer to \cite{Barnich:2009se,Barnich:2010eb,Barnich:2011ct,Kapec:2014opa,Kapec:2016jld,Strominger:2017zoo,Kapec:2017gsg,Laddha:2020kvp}.} More terrestrially, one can view celestial amplitude much in the same way as twistors, momentum twistors, 
scattering equations, which may help in illuminating hidden mathematical structures in quantum field theory that were not previously accessible from traditional calculations \cite{Elvang:2013cua}. 

In the last couple years, celestial amplitudes have garnered a lot of interest. Conformal primary wave function bases for various spins in different dimensions were constructed by Pasterski and Shao in \cite{Pasterski:2017kqt}.
Soft theorems were connected to the conserved currents on the celestial sphere in \cite{He:2015zea,Cheung:2016iub}. Explicit examples of tree level celestial amplitudes of gluons were computed in \cite{Pasterski:2017ylz, Schreiber:2017jsr}. Examples of scalar scattering have been shown in \cite{Lam:2017ofc,Banerjee:2017jeg,Nandan:2019jas}. Other variant maps have been constructed recently in \cite{Banerjee:2018gce,Banerjee:2018fgd,Banerjee:2019prz}. The investigation of the factorisation singularities of celestial amplitudes was done in \cite{Cardona:2017keg}.  Progress has also been made in the celestial four-point superstring amplitudes as well as graviton tree amplitudes \cite{Stieberger:2018edy, Puhm:2019zbl,Guevara:2019ypd} Recently, conformal soft theorems have been studied in \cite{Donnay:2018neh,Fan:2019emx,Banerjee:2019aoy,Pate:2019mfs,Adamo:2019ipt,Nandan:2019jas,Puhm:2019zbl,Guevara:2019ypd,Himwich:2019dug}. The authors in  \cite{Stieberger:2018onx} construct the generators of Poincar\'e and conformal groups in the celestial representation. Translating optical theorem in the conformal basis was addressed in \cite{Lam:2017ofc} and this work was followed by \cite{Nandan:2019jas} where conformal partial wave decomposition of celestial amplitudes was further discussed. The operator product expansion of the celestial sphere has been carried out in \cite{Ball:2019atb,Fotopoulos:2019tpe,Pate:2019lpp,Banerjee:2020kaa,Fotopoulos:2019vac}.

While scalar loops have been studied in \cite{Banerjee:2017jeg}, most construction of celestial amplitudes have occurred at tree-level.  In this work, we provide the first explicit construction of loop level celestial transform for external gluons and gravitons. More concretely, we focus on loop amplitudes where all external gluons and gravitons carry positive helicity (\emph{all-plus}) and the ones where all but one external particles carry positive helicity (\emph{one-minus}). These loop amplitudes are interesting for many reasons. It is a well known result that gluon as well as graviton amplitudes at tree level vanish for all-positive and one-minus external states \cite{Grisaru:1977px}. This statement can be proved using supersymmetric Ward identity, but nonetheless holds for quantum field theory with or without supersymmetry at tree-level.  However, the story of loops is different and very interesting: for supersymmetric field theories, such vanishing occurs for all-plus and one-minus amplitudes at any loop order; in contrast, for pure Yang-Mills and Einstein gravity, such amplitudes receive leading contributions at one loop (for gluons see \cite{Bern:2005hs,Mahlon:1993fe, Bern:1993qk} and gravitons \cite{Bern:1993sx,  Bern:1993wt, Dunbar:1994bn, Bern:1998sv, Bern:1998xc, Boels:2013bi}).\footnote{While we will restrict ourselves to one loop in this work, there has been a number of work on higher loops for the external states we are considering. Please see, \cite{Gehrmann:2015bfy, Dunbar:2016aux,Badger:2016ozq,Badger:2018enw,Bern:2015ooa,Dunbar:2016cxp,Badger:2019djh, Dalgleish:2020mof}.}Similarly, for one loop, one cannot construct such amplitudes using unitarity cuts in four dimensions as two particle cut leads to tree level expressions with at least one vanishing piece. Furthermore, these one loop amplitudes when integrated are relatively simple rational functions and contain no logarithmic divergences in four dimensions.\footnote{Moreover, these loop amplitudes have interesting factorization properties and have been studied using BCFW recursion relation \cite{Bern:2005hs, Farrow:2020voh, Kharel:2011vz} and using Berends-Giele type of recursion \cite{Mahlon:1993si} and more recently conformally invariant structure was investigated in \cite{Henn:2019mvc}}The simplicity and subtleties of these amplitudes make them ideal candidates to study spinning celestial amplitudes beyond tree level. 

We have organized this paper in the following way. In section~\ref{review} we present a detailed review of conformal primary wavefunctions and celestial amplitudes, which is followed by the rederivation of tree-level celestial amplitudes in our conventions. Thereafter, we switch to computation of loop level amplitudes in \secref{\ref{loop}} where we considered explicit four and five point results and discussed the structure of the higher point amplitudes. We end with future directions and conclusion, and collected  several technical details  in the appendix.
\section{Review}
\label{review}
\subsection{Conformal primary wavefunctions and celestial amplitudes}
We know that scattering amplitudes in $4d$ have the Poincar\'e symmetry $\ISO(1,3)$ which can be written as the semidirect product of translation and special orthogonal groups:\linebreak \mbox{$\ISO(1,3)=\mathrm{T}(1,3)\otimes_s \SO(1,3)$}. The standard momentum space enables us to work with irreducible representations of the translation group, which means translations act only by phases. $\SO(1,3)$ on the other hand acts in a quite complicated manner.

We can try to relate the $4d$ momentum space to some basis of  $\SO(1,3)$, or its universal covering group $\SL(2,\C)$. As $\SL(2,\C)$ is isomorphic to the global conformal group in $2d$, it is intriguing to expand the $4d$ amplitudes in terms of \emph{conformal primary wavefunctions}, objects that transforms covariantly as $2d$ conformal primary operators and satisfy relevant $4d$ equation of motion.

Before analyzing conformal primary wavefunctions in more detail, let us first set our notations. We will use the standard coordinates $z,\zb$ with $\zb=z^*$ to parametrize $\R^2$, and the $2d$ CFT lives at the compactification of this space, i.e. the Riemann sphere $\C_\infty$. We can then view this Riemann sphere as the boundary of a hyperbolic space $\mathrm{H}^3$ via AdS holography. We parametrize $\mathrm{H}^3$ with the coordinates $y_i$ with $i=0,1,2$ for $y_0>0$ where $\mathrm{H}^3$ metric is
\be 
ds^2_{\mathrm{H}^3}=\frac{dy_0^2+dy_1^2+dy_2^2}{y_0^2}
\ee 
We can then embed $\mathrm{H}^3$  as the upper branch of the unit hyperboloid in $\R^{1,3}$ for which we will use lightcone coordinates $x^\mu=(x^+,x^-,x^1,x^2)$ with the metric
\be 
g_{\mu\nu}=\begin{pmatrix}
0&\half&0&0\\\half&0&0&0\\0&0&1&0\\0&0&0&1
\end{pmatrix}
\ee 
where we define the lightcone coordinates in terms of the Cartesian ones as $x^\pm=x^3\pm x^0$. We then embed $y_i\in H^3$ in $y^\mu\in \R^{1,3}$ as
\be 
y^\mu\equiv\frac{1}{y_0}(1,-y_iy_i,y_1,y_2)
\ee 
where $y_iy_i\equiv y_0^2+y_1^2+y_2^2$ and where we see that $y_\mu y^\mu=-1$ as required.

Just as we embedded $y_i\rightarrow y^\mu$, we can embed $z\rightarrow x^\mu(z)$ because $d-$dimensional conformal groups can be parameterized with the null rays in $d+2$ dimensions.\footnote{The idea goes back to Dirac who realized that a conformal group in $\R^{p,q}$ dimensions being $\SO(q+1,q+1)$ can most naturally be described in the embedding $\R^{p+1,q+1}$ space \cite{Dirac:1936fq}.} In other words, \emph{the celestial sphere} can be parameterized in $\R^{1,3}$ as \mbox{$\{x^\mu(z)\in \R^{1,3}\;|\;x^\mu(z)\sim \lambda x^\mu(z),\lambda\in \R^+\}$} where we choose
\be 
\label{eq: null point}
x^\mu(z)\equiv2\left(1,-z\zb,\frac{z+\zb}{2},\frac{z-\zb}{2i}\right)
\ee  
for later convenience.

Below, we will write $y^\mu$ to denote a point in $\R^{1,3}$ constrained to lie on the upper branch of the unit hyperboloid, and $x^\mu(z)$ to denote a null vector in $\R^{1,3}$ whereas $x^\mu$ denotes any point in $\R^{1,3}$. In summary, $y_\mu y^\mu=-1$, $x_\mu(z)x^\mu(z)=0$ with $x^+(z)=2$, and $x_\mu x^\mu \in \R$.

With our notations set, we can view conformal primary wavefunctions as maps from $x^\mu\in\R^{1,3}$ to $z\in\C_\infty$, where these particular maps satisfy two conditions:
\begin{itemize}
\item They satisfy the equation of motion in $\R^{1,3}$
\item They transform as conformal primary operators under the action of $\SL(2,\C)$
\end{itemize}

A particularly transparent way to construct these objects for massive scalars can be roughly described as follows: we decompose the map \mbox{$\R^{1,3}\rightarrow\C_\infty$} into \mbox{$\R^{1,3}\rightarrow \mathrm{H}^3$} and \mbox{$\mathrm{H}^3\rightarrow \C_\infty$}, compose these maps, and integrate over whole $\mathrm{H}^3$. Indeed, the necessary ingredients for each map is relatively straightforward: the first map is simply a restriction to the paraboloid whereas second map is the familiar bulk to boundary propagator in $\mathrm{H}^3$. Hence, we can immediately define the \emph{massive scalar conformal primary wavefunction} $\phi^\pm_{\De,m}(x^\mu,z)$ as
\be 
\label{eq: massive conformal primary wavefunction}
\phi^\pm_{\De,m}(x^\mu,z)=\int\limits_{\mathrm{H}^3}[dy_i]G_{\De}(y_i,z)e^{\pm i m x_\mu y^\mu}
\ee 
Here, $e^{\pm i m x_\mu y^\mu}$ makes sure that EOM is satisfied, i.e. $\left(\partial_\mu\partial^\mu-m^2\right)\phi^\pm_{\De,m}(x^\mu,z)=0$, whereas $G_{\De}(y_i,z)$ ensures the correct transformation under $\SL(2,\C)$ action (note that $e^{\pm i m x_\mu y^\mu}$ is invariant under $\SL(2,\C)$). 

The closed form expression for $\phi^\pm_{\De,m}(x^\mu,z)$ in $\R^{1,d+1}$ reads as
\be 
\phi^\pm_{\De,m}(x^\mu,z)=\frac{2^{\frac{d+2}{2}}\pi^{\frac{d}{2}}}{(im)^{\frac{d}{2}}}\frac{\left(\sqrt{-x_\mu x^\mu}\right)^{\De-\frac{d}{2}}}{\left(-x_\mu x^\mu(z)\mp i\epsilon\right)^\De}K_{\De-\frac{d}{2}}\left(m\sqrt{x_\mu x^\mu}\right)
\ee 
One can similarly write down the \emph{massless} spin-0,1,2 conformal primary wavefunctions $\phi^\pm_{\De}(x^\mu,z)$, $A^{\De,\pm}_{\mu,a}(x^\mu,z)$ and $h^{\De,\pm}_{\mu_1,\mu_2,a_1,a_2}(x_\mu,z)$  as
\bea [eq: massless conformal primary wavefunctions]
\phi^\pm_{\De}(x^\mu,z)=&\frac{(\mp i)^\De\Gamma(\De)}{\left(-x_\mu x^\mu(z)\mp i\epsilon\right)^{\De}}\label{eq: spin-0 massless conformal primary wavefunctions}\\
A^{\De,\pm}_{\mu,a}(x^\mu,z)=&-\frac{1}{\left(-x_\mu x^\mu(z)\mp i\epsilon\right)^{\De-1}}T_{\mu,a}^\pm(x^\mu,z)
\\
h^{\De,\pm}_{\mu_1,\mu_2,a_1,a_2}(x_\mu,z)=&\frac{\delta^{b_1}_{\;\;(a_1}\delta^{b_2}_{\;\;a_2)}-\frac{1}{d}\delta_{a_1a_2}\delta^{b_1b_2}}{\left(-x_\mu x^\mu(z)\mp i\epsilon\right)^{\De-2}}T_{\mu_1,b_1}^\pm(x^\mu,z,\zb)T_{\mu_2,b_2}^\pm(x^\mu,z)
\eea 
for
\be 
T_{\mu,a}^\pm(x^\mu,z)\equiv \frac{\partial}{\partial x^{\mu}}\frac{\partial}{\partial z_{a}}\log\left(-x_\mu x^\mu(z)\mp i\epsilon\right)
\ee 
where $z_a\equiv z_{1,2}$ for $z_1\equiv\frac{z+\zb}{2}$ and $z_2\equiv\frac{z-\zb}{2i}$.

Conformal primary wavefunctions, as interesting as they may be, would not be so much of usage if they did not form a complete basis for on shell wavefunctions in $\R^{1,d+1}$. Indeed, as shown in \cite{Pasterski:2017kqt}, we can identify on-shell momenta $p_\mu$ as $p_\mu=m y_\mu$ which satisfies $ p_\mu  p^\mu=-m^2$ as required. With this identification, we can rewrite  \equref{eq: massive conformal primary wavefunction}
\be 
\label{eq: basis change for massive scalar}
\phi^\pm_{\De,m}(x^\mu,z,\zb)=\int\limits_{\mathrm{H}^3}[dy_i]G_{\De}(y_i,z,\zb)e^{\pm i x_\mu  p^\mu}
\ee 
which can be seen as a basis transformation from momentum space spanned by $\{e^{\pm i x_\mu  p^\mu}\}$ to a new basis spanned by $\phi^\pm_{\De,m}(x^\mu,z)$. The original basis was labeled with \mbox{$\{ p_\mu\in\R^{1,3}| p_\mu p^\mu=-m^2\}$} whereas the new basis is labeled with $\{\left(\De,z\right)\in\left(\C,\C_\infty\right)|\De=1+i\R\}$.\footnote{The restriction of $\De$ to \emph{principal series}, i.e. $\De=\frac{d}{2}+i\R$, is a necessary condition for the conformal quadratic Casimir operator to be self-adjoint, which ensures by the spectral theorem that it has an orthonormal basis of eigenvectors; thus $\phi^\pm_{1+i\R,m}$ form an orthonormal basis. For $m\ne 0$, shadow symmetry\footnotemark \; makes $\frac{d}{2}\pm i\nu$ linearly dependent, hence we only take half of the principal series, i.e. $\De=\frac{d}{2}+i\R_{\ge 0}$.}\footnotetext{Shadow transformation is an interwining map from an operator in representation $(\De,\rho)$ to another operator in representation $(d-\De,\rho^R)$ where $\rho$ is an $\SO(d)$ irrep and where $\rho^R$ denotes the reflected representation. In odd dimensions, one can take $\rho^R\simeq\rho$ hence shadow transformation amounts to $\De\rightarrow d-\De$ which relates the principal series representations $\De=\frac{d}{2}\pm i \nu$.}

The basis transformation for massless scalars is far more intuitive. To see that, we first write the momentum vector as 
\be 
p^\mu=\epsilon\omega x^\mu(z)
\ee 
where $\epsilon=1(-1)$ for outgoing (incoming) momentum. We can interpret $z$ coordinate trivially: it just parametrizes the \emph{direction} of the momentum vector on the celestial sphere. Now, we only need to relate the \emph{magnitude} of the momentum, $\omega$, to the scaling dimension of the conformal primary wavefunction, i.e. $\Delta$. We can see this relation
between the basis vectors in the form of a Mellin transformation
\be  
\label{eq: basis change for massless scalar}
\phi^\pm_{\De}(x^\mu,z,\zb)=\int\limits_0^\infty d\omega \omega^{\De-1}e^{\pm i\left(x_\mu  p^\mu \pm i\epsilon\right)}
\ee 
which follows from \equref{eq: spin-0 massless conformal primary wavefunctions} and $ p^\mu=\omega x^\mu(z)$. Physically, this means that the Lorentz boosts, which acts as $x_\mu  p^\mu \rightarrow \lambda x_\mu  p^\mu $, become dilatation in conformal primary wavefunction bases, i.e. $\phi^\pm_{\De}(x^\mu,z)\rightarrow \lambda^{-\De}\phi^\pm_{\De}(x^\mu,z)$. 

We can see that the relation between plane waves and massless conformal primary wavefunctions is implemented by a Mellin transformation for spin 1 and 2 cases as well, though there are subtleties of gauge and diffeomorphism invariances.\footnote{For example, $A^{\De,\pm}_{\mu,a}(x^\mu,z,\zb)$ can be related to plane waves with a Mellin transformation only if $\De\ne 1$ in $d=4$. This follows from the fact that $A^{1,\pm}_{\mu,a}(x^\mu,z,\zb)$ is simply \emph{a pure gauge term} in $d=4$ hence it cannot be related to the physical plane wave solution.} We refer the reader to \cite{Pasterski:2017kqt} for further details.

With the bases of conformal primary wavefunctions set up, we can now construct \emph{celestial amplitudes}. The traditional amplitudes with external particles being momentum eigenstates can be cast into the form
\be 
\label{eq: traditional amplitude as fourier transform}
\cA(p_1,\dots,p_n)=\int \prod\limits_{i=1}^nd^4x_i e^{i(p_i)_\mu x_i^\mu} a(x_1,\dots x_n)
\ee 
where $p_i$ are outgoing momenta of external scalars and where $a(x_1,\dots x_n)$ is the rest of the amplitude. The celestial amplitude $\tl\cA$ requires all external wavefunctions to be conformal primary wavefunctions instead, hence 
\be 
\label{eq: celestial amplitude as position integration}
\tl\cA^{\De_1,\dots,\De_n}(z_1,\dots,z_n)=\int \prod\limits_{i=1}^nd^4x_i \phi^+_{\De_i}(x_i^\mu,z_i) a(x_1,\dots x_n)
\ee 
For example, for cubic vertex interaction $\cL\sim \lambda\phi_1\phi_2\phi_3$, we can compare the three-point scattering amplitudes as follows
\bea
\cA(p_1,p_2,p_3)=&\int \prod\limits_{i=1}^3d^4x_i e^{i(p_i)_\mu x_i^\mu} (i\lambda)=i\lambda(2\pi)^4\delta^4(\vec{p}_1+\vec{p}_2+\vec{p}_3)
\\
\tl\cA^{\De_1,\De_2,\De_3}(z_1,z_2,z_3)=&\int \prod\limits_{i=1}^3d^4x_i \phi^+_{\De_i}(x_i^\mu,z_i) (i\lambda)\sim \frac{\lambda}{\abs{z_{12}}^{\De_{123}}\abs{z_{23}}^{\De_{231}}\abs{z_{31}}^{\De_{312}}}
\eea 
for $z_{ij}\equiv z_i-z_j$ and $\De_{ijk}\equiv \De_i+\De_j-\De_k$. As expected by the conformal covariance of $\tl\cA$, the three point celestial amplitude in $4d$ takes the form of the $3-$point CFT correlator in $2d$.

By using the basis change in \equref{eq: basis change for massive scalar} for massive scalars or \equref{eq: basis change for massless scalar} for massless scalars, we can relate $\cA(p_i)$ and $\tl\cA^{\De_i}(z_i)$. In fact, for all massless spin-0,1,2 external particles, the transition from $\cA$ to $\tl \cA$ takes the form of a Mellin transformation, which implements the map from momentum eigenstates to boost eigenstates:
\be 
\label{eq: Mellin transform}
\tl\cA_{J_1\dots J_n }^{\De_1,\dots,\De_n}(z_1,\dots,z_n)=
\left(\prod\limits_{i=1}^n\int\limits_0^\infty d\omega_i\omega_i^{\De_i-1}\right)
\cA_{ j_1\dots j_n}(\omega_1,\dots,\omega_n;z_1,\dots,z_n)
\ee 
where we used $ p^\mu=\omega x^\mu(z,\zb)=\omega\left(2,-2z\zb,z+\zb,-i(z-\zb)\right)$ and where $2d$ spin $J_i$ is identified with $4d$ helicity $j_i$: $J_i=j_i$.

$n-$successive Mellin transforms are relatively straightforward, however we can simplify it further via the exploitation of  the covariance of $\tl\cA$ under boosts (dilations in the celestial sphere) by switching to simplex variables
\be 
s\equiv \sum\limits_{i=1}^n\omega_i\;,\quad \sigma_i\equiv\frac{\omega_i}{s}
\ee
under which \equref{eq: Mellin transform} becomes
\begin{multline}
\label{eq: celestial amplitude from simplex coordinates}
\tl\cA_{J_1\dots J_n }^{\De_1,\dots,\De_n}(z_1,\dots,z_n)=
2\pi\delta\left(i(\kappa-n)+\sum\limits_{i=1}^n\lambda_i\right)
\prod\limits_{k=1}^n\left(\int\limits_0^1d\sigma_k\sigma_k^{i\lambda_k}\right) \delta^4\left(\sum\limits_{i=1}^nq_i^{\mu}\sigma_i\right)
\\\x\delta\left(\sum\limits_{i=1}^n\sigma_i-1\right)\Amp{}{j_1\dots j_n}{\s_1\dots\s_n}{z_1\dots z_n}
\end{multline}
where we wrote down $\De$ on the principal series as
\be 
\De=1+i\lambda\;,\quad \lambda\in\R
\ee 
and where we define the stripped amplitude $\Amp{}{j_1\dots j_n}{\s_1\dots\s_n}{z_1\dots z_n}$ as 
\be 
\label{eq: definition of amp}
\cA_{ j_1\dots j_n}(\omega_1,\dots,\omega_n;z_i,\dots,z_n)
=s^{-\kappa}
\Amp{}{j_1\dots j_n}{\s_1\dots\s_n}{z_1\dots z_n} \delta^4\left(\sum\limits_{i=1}^nq_i^{\mu}\sigma_i\right)
\ee 
Here $\kappa$ is the overall momentum scaling of the amplitude\footnote{For example, for tree level MHV amplitudes, $\kappa=n$ as we can easily see from \equref{eq: tree level MHV amplitude in celestial coordinates}.}, i.e. 
\be 
\label{eq: definition of kappa}
\cA_{ j_1\dots j_n}(\Lambda \omega_1,\dots,\Lambda \omega_n;z_i,\dots,z_n)
=\Lambda^{-\kappa}
\cA_{ j_1\dots j_n}(\omega_1,\dots,\omega_n;z_i,\dots,z_n)
\ee 
and we defined
\be 
q_i^{\mu}\equiv  \epsilon_ix^\mu(z_i)
\ee 
for brevity.

One can leverage the covariance of celestial amplitudes under the conformal group by going to a \emph{conformal frame} where we choose\footnote{Given any three points, we can first use translations to fix $z_1=0$, then special conformal transformation to take $z_2\rightarrow \infty$, then dilation to bring $z_3$ to unit circle, and finally rotation to get $z_3=1$. As this exhaust all conformal transformations, $z_{n>3}$ remains unfixed. By applying these transformations in reverse, we can get any amplitude $\tl\cA(z_1,z_2,z_3,z_4,\dots,z_n)$ from $\tl\cA(0,\infty,1,z_4',\dots,z_n')$. See Appendix~\ref{sec: conformal frame} for further details.}
\be 
z_1=0\;,\quad z_2=\infty\;,\quad z_3=1
\ee 
As dilation acts inversely at infinity, a correct procedure to put an operator at $z=\infty$ is by the limit
\be 
\label{eq: putting an operator at infinity}
\cO(\infty)=\lim\limits_{L\rightarrow\infty}L^{2\De_\cO}\cO(L)
\ee 
hence we define the celestial amplitude in this conformal frame as\footnote{
We use 
\scriptsize
\begin{multline}
\lim\limits_{L\rightarrow\infty}\prod\limits_{k=1}^n\left(\int\limits_0^1d\sigma_k\right)
\delta^4\left(\sum\limits_{i=1}^nq_i^{\mu}\sigma_i\right)
\delta\left(\sum\limits_{i=1}^n\sigma_i-1\right)
f(\s_1,\s_2,\s_3,\dots,\s_n;z_i)=
\prod\limits_{k=3}^n\left(\int\limits_0^1d\sigma_k\right)
\delta\left(1+\sum\limits_{i=3}^n\left(\epsilon_1\epsilon_i-1\right) \sigma_i\right)
\\\x
\delta\left(\sum\limits_{i=3}^n\epsilon_i \sigma_iz_i\right)
\delta\left(\sum\limits_{i=3}^n\epsilon_i \sigma_i\zb_i\right)
\lim\limits_{L\rightarrow\infty}\frac{1}{4L^2}
f\left(1-\sum\limits_{i=3}^n\s_i,-\frac{1}{L^2}\sum\limits_{i=3}^n\epsilon_2\epsilon_i \sigma_i\left(1+z_i\zb_i\right),\s_3,\dots,\s_n;z_i\right)
\end{multline}
\normalsize
which follows from the conformal frame we choose and $q_i^\mu=\epsilon_ix^\mu(z_i)$ with our choice of $x^\mu(z)$ in \equref{eq: null point}.
}
\small
\begin{multline}
\label{eq: celestial amplitude from simplex coordinates 2}
\tl\cA_{J_1\dots J_n }^{\De_1,\dots,\De_n}(0,\infty,1,z_4,\dots,z_n)=
\frac{\pi}{2}\delta\left(i(\kappa-n)+\sum\limits_{i=1}^n\lambda_i\right)
\prod\limits_{k=3}^n\left(\int\limits_0^1d\sigma_k\sigma_k^{i\lambda_k}\right)
\left(1-\sum\limits_{i=3}^n\s_i\right)^{i\lambda_1}
\\\x 
\left(-\sum\limits_{i=3}^n\epsilon_2\epsilon_i \sigma_i\left(1+z_i\zb_i\right)\right)^{i\lambda_2}
\delta\left(\sum\limits_{i=3}^n\epsilon_i \sigma_iz_i\right)
\delta\left(\sum\limits_{i=3}^n\epsilon_i \sigma_i\zb_i\right)
\\\x 
\delta\left(1+\sum\limits_{i=3}^n\left(\epsilon_1\epsilon_i-1\right) \sigma_i\right)
\lim\limits_{L\rightarrow\infty} 
\Amp{}{j_1\dots j_n}{1-\sum\limits_{i=3}^n\s_i,\;-L^{-2}\sum\limits_{i=3}^n\epsilon_2\epsilon_i \sigma_i\left(1+z_i\zb_i\right),\;\s_3,\;\dots,\;\s_n}{0,\;L,\;1,\; z_4,\;\dots,\; z_n}
\end{multline}
\normalsize
where delta functions of momentum conservation along the lightcone coordinates were immediately employed to remove $\s_{1,2}$ integrations via the use of \equref{eq: null point} in \equref{eq: celestial amplitude from simplex coordinates}, hence the delta functions above are due to the momentum conservation along transverse directions and due to the normalization condition of the simplex variables, i.e. $\sum\limits_{i=1}^n\s_i=1$.

By using \equref{eq: delta rearrangement}, we can rewrite this equation as 
\begin{subequations}
\label{eq: most general results in conformal frame}
\small
\begin{multline}
\label{eq: celestial amplitudes final form 1}
\tl\cA_{J_1\dots J_n }^{\De_1,\dots,\De_n}(0,\infty,1,z_4,\dots,z_n)=
\frac{\pi}{2}\frac{\mathcal{U}(\b_i)}{\abs{M_{1,2,3}}}\delta\left(i(\kappa-n)+\sum\limits_{i=1}^n\lambda_i\right)
\prod\limits_{k=6}^n\left(\int\limits_0^1d\sigma_k\sigma_k^{i\lambda_k}\right)\b_1^{i\lambda_3}\b_2^{i\lambda_4}\b_3^{i\lambda_5}
\\\x 
\left(1-\sum\limits_{i=1}^3\b_i-\sum\limits_{i=6}^n\s_i\right)^{i\lambda_1}
\left(-\sum\limits_{i=1}^3\epsilon_2\epsilon_{i+2} \b_i\left(1+z_{i+2}\zb_{i+2}\right)-\sum\limits_{i=6}^n\epsilon_2\epsilon_i \sigma_i\left(1+z_i\zb_i\right)\right)^{i\lambda_2}
\\\x 
\lim\limits_{L\rightarrow\infty} 
\Amp{}{j_1\dots j_n}{1-\sum\limits_{i=1}^3\b_i-\sum\limits_{i=6}^n\s_i,\;-L^{-2}\left(\sum\limits_{i=1}^3\epsilon_2\epsilon_{i+2} \b_i\left(1+z_{i+2}\zb_{i+2}\right)+\sum\limits_{i=6}^n\epsilon_2\epsilon_i \sigma_i\left(1+z_i\zb_i\right)\right),\;\b_1,\;\b_2,\;\b_3,\;\s_6,\;\dots,\;\s_n}{0,\;L,\;1,\; z_4,\;\dots,\; z_n}
\end{multline}
\normalsize
for $n>4$; in particular, we do not have any integrals left for $n=5$:
\small
\begin{multline}
\label{eq: celestial amplitudes final form 2}
\tl\cA_{J_1\dots J_5 }^{\De_1,\dots,\De_5}(0,\infty,1,z_4,z_5)=
\frac{\pi}{2}\frac{\mathcal{U}(\b_i)}{\abs{M_{1,2,3}}}\delta\left(i(\kappa-5)+\sum\limits_{i=1}^5\lambda_i\right)
\b_1^{i\lambda_3}\b_2^{i\lambda_4}\b_3^{i\lambda_5}
\left(1-\sum\limits_{i=1}^3\b_i\right)^{i\lambda_1}
\\\x 
\left(-\sum\limits_{i=1}^3\epsilon_2\epsilon_{i+2} \b_i\left(1+z_{i+2}\zb_{i+2}\right)\right)^{i\lambda_2}
\lim\limits_{L\rightarrow\infty} 
\Amp{}{j_1\dots j_n}{1-\sum\limits_{i=1}^3\b_i,\;-L^{-2}\sum\limits_{i=1}^3\epsilon_2\epsilon_{i+2} \b_{i}\left(1+z_{i+2}\zb_{i+2}\right),\;\b_1,\;\b_2,\;\b_3}{0,\;L,\;1,\; z_4,\;z_5}
\end{multline}
\normalsize

For $n=4$, we instead use \equref{eq: delta rearrangement 2}, with which \equref{eq: celestial amplitude from simplex coordinates 2} becomes
\small
\begin{multline}
\label{eq: celestial amplitudes final form 3}
\tl\cA_{J_1\dots J_4 }^{\De_1,\dots,\De_4}(0,\infty,1,z_4)=
\frac{\pi}{2}\mathcal{U}(\b_i)\delta\left(\zb_4-z_4\right)\delta\left(i(\kappa-4)+\sum\limits_{i=1}^4\lambda_i\right)
\b_1^{i\lambda_3}\b_2^{i\lambda_4}
\left(1-\sum\limits_{i=1}^2\b_i\right)^{i\lambda_1}
\\\x 
\left(-\sum\limits_{i=1}^2\epsilon_2\epsilon_{i+2} \b_i\left(1+z_{i+2}\zb_{i+2}\right)\right)^{i\lambda_2} 
\lim\limits_{L\rightarrow\infty} 
\Amp{}{j_1\dots j_4}{1-\sum\limits_{i=1}^2\b_i,\;-L^{-2}\sum\limits_{i=1}^2\epsilon_2\epsilon_{i+2} \b_i\left(1+z_{i+2}\zb_{i+2}\right),\;\b_1,\;\b_2}{0,\;L,\;1,\; z_4}
\end{multline}
\normalsize
\end{subequations}
For details regarding the coefficients $\b$, $M_{1,2,3}$, and the function $\mathcal{U}$ in \equref{eq: most general results in conformal frame}, please see Appendix~\ref{sec: generalized Cramer's rule}. One curious observation regarding \equref{eq: most general results in conformal frame} is that the celestial amplitudes are on the principal series (i.e. $\lambda\in\R$) only if $\ka=n$; in other words, we need to analytically continue off the principal series if the amplitude $\cA_{j_1\dots j_n}(p_i)$ does not have the mass dimension $-n$. The celestial amplitude $\tl\cA$ being off the principal series means that the CFT operators are no longer in the unitary representation of the group\footnote{One should not confuse the unitarity of the group representation, which has to do with the self-adjointness of the Casimir operator, with the unitarity of the field theory, which is the requirement that norms of the states in Hilbert space are non-negative. Indeed, unitarity of the CFT actually requires other conditions for $\De$ than it being on the principal series (i.e. $\De=1+i\R$); for example, we need $\De\ge l+d-1-\frac{d}{2}\delta_{l,0}$ for a CFT$_d$ operator in symmetric traceless tensor representation for the CFT to be unitary.}; and in particular, it means that the conformal primary wavefunctions would not constitute an orthonormal basis for these amplitudes.\footnote{ Principal series representations are actually not the only unitary representations for conformal groups, but they are the only tempered  unitary representation that appears in $2d$ \cite{Gadde:2017sjg}.} Nevertheless, the procedure of doing harmonic analysis for the CFT correlators on the principal series and then analytically continuing them to the regions of interest is relatively well-known and has been extensively used to extract CFT data through Euclidean inversion formula \cite{Kravchuk:2018htv,Liu:2018jhs,Karateev:2018oml,Albayrak:2020rxh}.\footnote{For possible subtleties regarding the analytic continuation, see \cite{Simmons-Duffin:2017nub} and references therein.} We should also note that \emph{the generalized soft limit} on celestial sphere may relate amplitudes on principal series to amplitudes off principal series as we have
\begin{subequations}
\label{eq: generalized soft limit}
\be
\lim\limits_{\lambda_m\rightarrow 0}\lambda_m\tl\cA_{J_1\dots J_n }^{\De_1,\dots,\De_n}(z_1,\dots,z_n)
=-2i \sum\limits_{a}f^{\text{soft}}_{a,J_m}(z_i)
\tl\cA_{J_1,\dots,J_{m-1},J_{m+1},\dots,J_n}^{\De_1^{(a)},\dots,\De^{(a)}_{m-1},\De^{(a)}_{m+1},\dots,\De^{(a)}_n}(z_1,\dots,z_{m-1},z_{m+1},\dots,z_n)
\ee
for
\be 
\De_k=1+i \lambda_k\;,\quad \De_k^{(a)}=1+i \lambda_k+n_k^{(a)}
\ee 
\end{subequations}
thus in the rest of the paper we will not dwell on the appearance of  $\delta\left(i(\kappa-n)+\sum\limits_{i=1}^n\lambda_i\right)$ in \equref{eq: most general results in conformal frame}.\footnote{We can  take the soft limit in momentum space as
\begin{multline}
\lim\limits_{\w_m\rightarrow 0}\w_m \cA_{J_1\dots J_n }(\w_1,\dots,\w_n;z_1,\dots,z_n)
=\left(\sum\limits_af^{\text{soft}}_{a,J_m}(z_i)\prod\limits_{k=1,\dots,m-1,m+1,\dots,n}(\w_k)^{n_k^{(a)}} \right)
\\\x\cA_{J_1\dots J_{m-1},J_{m+1},\dots,J_n }(\w_1,\dots,\w_{m-1},\w_{m+1},\dots,\w_n;z_1,\dots,z_{m-1},z_{m+1},\dots,z_n)
\end{multline}
for the soft factor $f^{\text{soft}}_{a,J_m}(z_i)$ and the coefficients $n_k^{(a)}$ which depend on the amplitude under consideration. We can then derive \equref{eq: generalized soft limit} with this equation and the representation of delta distribution as
\be 
\delta(x)=\frac{i}{2}\lim\limits_{\lambda_m\rightarrow 0}\lambda_m\w_m^{i\lambda_m -1}
\ee 
For example, in the case of  n point graviton amplitude with the choice of $m=n$, we have $f^{\text{soft}}_{a,J_n=+}(z_i)=\frac{1}{\e_{a}\e_n}\frac{\zb_{na}z_{xa}z_{ya}}{z_{na}z_{xn}z_{yn}}$ and $n^{(a)}_k=\delta^a_k$ for \mbox{$a=1,\dots,n-1$} where  $x$ and $y$ are properly chosen reference points \cite{Puhm:2019zbl}.
}

\subsection{Tree level gluon celestial amplitudes}
In this section, we will review some tree-level results before we move on to computing loop-level gluon celestial amplitudes in the body.

One of the simplest tree-level example that we can consider is
color-ordered MHV  amplitude, which in spinor helicity notation takes the form
\be 
\cA^{\texttt{MHV}}_{--+\dots +}(p_i)=\frac{\<12\>^3}{\<23\>\dots\<n1\>}\delta^4\left(\sum\limits_{i=1}^np_i^\mu\right)
\ee 
where we are following the notation of \cite{Pasterski:2017ylz} for spinor helicity formalism. In particular,
\be 
\label{eq: spinor helicity convention}
\left[ij\right]=2\sqrt{\omega_i\omega_j}\zb_{ij}\;,\quad \<ij\>=-2\epsilon_i\epsilon_j\sqrt{\omega_i\omega_j}z_{ij}
\ee 
hence 
\be 
\label{eq: tree level MHV amplitude in celestial coordinates}
\cA^{\texttt{MHV}}_{--+\dots +}(\omega_1,\dots,\w_n;z_1,\dots, z_n)=(-2)^{4-n}\frac{z_{12}^4}{z_{12}z_{23}\dots z_{n1}}
\frac{\omega_1^2\omega_2^2}{\omega_1\omega_2\dots\omega_n}
\delta^4\left(\sum\limits_{i=1}^n\epsilon_i\omega_ix^\mu(z_i)\right)
\ee 
Clearly, $\kappa=n$ for this amplitude and we can write down
\be 
\Amp{\texttt{MHV}}{--+\dots +}{\s_1\dots\s_n}{z_1\dots z_n}=(-2)^{4-n}\frac{z_{12}^4}{z_{12}z_{23}\dots z_{n1}}
\frac{\sigma_1^2\sigma_2^2}{\sigma_1\sigma_2\dots\sigma_n}
\ee 
for which we have
\small 
\begin{multline}
\lim\limits_{L\rightarrow\infty} 
\Amp{\texttt{MHV}}{--+\cdots +}{1-\sum\limits_{i=1}^3\b_i-\sum\limits_{i=6}^n\s_i,\;-L^{-2}\left(\sum\limits_{i=1}^3\epsilon_2\epsilon_{i+2} \b_i\left(1+z_{i+2}\zb_{i+2}\right)+\sum\limits_{i=6}^n\epsilon_2\epsilon_i \sigma_i\left(1+z_i\zb_i\right)\right),\;\b_1,\;\b_2,\;\b_3,\;\s_6,\;\dots,\;\s_n}{0,\;L,\;1,\; z_4,\;\dots,\; z_n}
\\
=(-2)^{4-n}\frac{1}{(1-z_{4})z_{45}z_{56}\dots z_{(n-1)n}z_n}\frac{1}{\b_1\b_2\b_3}
\frac{1}{\sigma_6\dots\sigma_n}\left(1-\sum\limits_{i=1}^3\b_i-\sum\limits_{i=6}^n\s_i\right)
\\\x
\left(\sum\limits_{i=1}^3\epsilon_2\epsilon_{i+2} \b_i\left(1+z_{i+2}\zb_{i+2}\right)+\sum\limits_{i=6}^n\epsilon_2\epsilon_i \sigma_i\left(1+z_i\zb_i\right)\right)
\end{multline}
\normalsize
Therefore, \equref{eq: celestial amplitudes final form 1} becomes
\begin{multline}
\left(\tl\cA^{\texttt{MHV}}\right)_{--+\cdots + }^{\De_1,\dots,\De_n}(0,\infty,1,z_4,\dots,z_n)=\frac{\pi(-2)^{3-n}}{(1-z_{4})z_{45}z_{56}\dots z_{(n-1)n}z_n}\frac{\mathcal{U}(\b_i)}{\abs{M_{1,2,3}}}\delta\left(\sum\limits_{i=1}^n\lambda_i\right)
\prod\limits_{k=6}^n\left(\int\limits_0^1d\sigma_k\sigma_k^{i\lambda_k-1}\right)
\\\x 
\b_1^{i\lambda_3-1}\b_2^{i\lambda_4-1}\b_3^{i\lambda_5-1}
\left(1-\sum\limits_{i=1}^3\b_i-\sum\limits_{i=6}^n\s_i\right)^{i\lambda_1+1}
\left(-\sum\limits_{i=1}^3\epsilon_2\epsilon_{i+2} \b_i\left(1+z_{i+2}\zb_{i+2}\right)-\sum\limits_{i=6}^n\epsilon_2\epsilon_i \sigma_i\left(1+z_i\zb_i\right)\right)^{i\lambda_2+1}
\end{multline}
where $n=5$ case can be straightforwardly written as 
\begin{multline}
\left(\tl\cA^{\texttt{MHV}}\right)_{--+\cdots + }^{\De_1,\dots,\De_n}(0,\infty,1,z_4,z_5)=\frac{\pi}{4 (1-z_{4})z_{45}z_5}\frac{\mathcal{U}(\b_i)}{\abs{M_{1,2,3}}}\delta\left(\sum\limits_{i=1}^5\lambda_i\right)
\b_1^{i\lambda_3-1}\b_2^{i\lambda_4-1}\b_3^{i\lambda_5-1}
\\\x 
\left(1-\sum\limits_{i=1}^3\b_i\right)^{i\lambda_1+1}
\left(-\sum\limits_{i=1}^3\epsilon_2\epsilon_{i+2} \b_i\left(1+z_{i+2}\zb_{i+2}\right)\right)^{i\lambda_2+1}
\end{multline}

By using the prescription detailed in Appendix~\ref{sec: generalized Cramer's rule}, we can compute $\b_i$ and write down the explicit expression for any given momenta; for example, we have
\begin{multline}
\left(\tl\cA^{\texttt{MHV}}\right)_{--+\cdots + }^{\De_1,\dots,\De_n}(0,\infty,1,z_4,\dots,z_n)\evaluated_{\substack{p_{2,3}\text{ : incoming}\\p_{1,k\ge 4}\text{ : outgoing}}}=\frac{\pi(-2)^{3-n}}{(1-z_{4})z_{45}z_{56}\dots z_{(n-1)n}z_n}\frac{\mathcal{U}(\b_i)}{\abs{\chi_{45}}}\delta\left(\sum\limits_{i=1}^n\lambda_i\right)
\\\x
\prod\limits_{k=6}^n\left(\int\limits_0^1d\sigma_k\sigma_k^{i\lambda_k-1}\right)\left(\frac{\chi_{34}+\chi_{53}+\chi_{45}\left(1-2\sum\limits_{k=6}^n\s_k\right)-2\sum\limits_{k=6}^n(\chi_{4k}-\chi_{5k})\s_k}{\chi_{45}}\right)^{i\lambda_1+1}
\\\x
\left(\frac{\chi_{34}\chi_5+\chi_{45}\left(\chi_3-2\sum\limits_{k=6}^n\chi_{k}\s_k\right)+\chi_{53}\chi_4}{\chi_{54}}\right)^{i\lambda_2+1}
\left(\frac{\chi_{53}+2\sum\limits_{k=6}^n\chi_{5k}\s_k}{\chi_{54}}\right)^{i\lambda_4-1}\left(\frac{\chi_{43}+2\sum\limits_{k=6}^n\chi_{4k}\s_k}{\chi_{45}}\right)^{i\lambda_5-1}
\end{multline}
for
\be 
\chi_i\equiv 1+z_i\zb_i\;,\quad \chi_{ij}\equiv z_i\zb_j-z_j\zb_i\quad\text{ for }z_3=\zb_3=1
\ee 
where $\mathcal{U}(\b_i)=0,1$ and it should be understood as a reminder that the expression is nonzero only for certain regions of $z_i$, regions whose explicit description we will not provide for the most generic case.

For $n=5$, the expression significantly simplifies; in particular,
\begin{multline}
\left(\tl\cA^{\texttt{MHV}}\right)_{--+++ }^{\De_1,\dots,\De_n}(0,\infty,1,z_4,z_5)\evaluated_{\substack{p_{2,3}\text{ : incoming}\\p_{1,4,5}\text{ : outgoing}}}=-\frac{\pi \mathcal{U}(\b_i)}{4(1-z_{4})z_{45}z_{5}}\delta\left(\sum\limits_{i=1}^5\lambda_i\right)
\\\x
\left(\chi_{43}+\chi_{35}+\chi_{54}\right)^{i\lambda_1+1}
\left(\chi_{34}\chi_5+\chi_{45}\chi_3+\chi_{53}\chi_4\right)^{i\lambda_2+1}\left(\chi_{54}\right)^{i\lambda_3-1}
\left(\chi_{53}\right)^{i\lambda_4-1}\left(\chi_{34}\right)^{i\lambda_5-1}
\end{multline}
if we restrict to $\chi_{54}\in \R^+$.

Of course, there is nothing specific about choosing second and third momenta to be outgoing and the rest incoming; for example,
\begin{multline}
\left(\tl\cA^{\texttt{MHV}}\right)_{--+++ }^{\De_1,\dots,\De_n}(0,\infty,1,z_4,z_5)\evaluated_{\substack{p_{4,5}\text{ : incoming}\\p_{1,2,3}\text{ : outgoing}}}=-\frac{\pi \mathcal{U}(\b_i)}{4(1-z_{4})z_{45}z_{5}}\delta\left(\sum\limits_{i=1}^5\lambda_i\right)
\\\x
\left(-\chi_{43}-\chi_{35}-\chi_{54}\right)^{i\lambda_1+1}
\left(\chi_{34}\chi_5+\chi_{45}\chi_3+\chi_{53}\chi_4\right)^{i\lambda_2+1}\left(\chi_{54}\right)^{i\lambda_3-1}
\left(\chi_{53}\right)^{i\lambda_4-1}\left(\chi_{34}\right)^{i\lambda_5-1}
\end{multline}
if we restrict to $\chi_{53}+\chi_{34}\in \R^+$.

\section{Loop amplitudes on the celestial sphere}
\label{loop}
In this section, we will consider gluon and graviton loop amplitudes with all-plus helicity (all external particles have positive helicity) and one-minus helicity (all but one external particles have positive helicity). These helicity configurations are particularly interesting choices as their tree-level counterparts vanish as shown in  \cite{Grisaru:1977px}. These \emph{rational} amplitudes are also interesting for other reasons; for instance, they are not cut-constructible through unitarity cuts in four dimensions. In addition their expressions take surprisingly compact forms, reminiscent of tree level amplitudes. Finally, they are free of logarithmic divergences, which make them ideal candidates for celestial amplitudes beyond tree level.
\subsection{Four point amplitudes}
\label{sec: four point loop}
\begin{table}
	\centering
	\caption{\label{eq: support function for 4 point amplitudes} The break-down of the support of the four point celestial amplitude on $\R$ depending on which momenta lie on future lightcone ($\e=1$) and which momenta lie on past lightcone ($\e=-1$).}
\begin{tabular}{cc}
\hline\hline
\textbf{Case} & \textbf{Region }$\mathcal{U}(\b_i)$\textbf{ is }$1$
\\\hline
$\e_1=\e_4=-\e_3$ & $z_4>\half$
\\
$\e_1=\e_3=-\e_4$ & $2\ge z_4 \ge 0$
\\
$\e_3=\e_4=-\e_1$ & $0>z_4$
\\\hline\hline
\end{tabular}
\end{table} 
Let us start with the four-point celestial amplitudes for gluons and gravitons.
We have seen that a generic four point  celestial amplitude can be computed from \equref{eq: celestial amplitudes final form 3}, which becomes
\begin{multline}
\label{eq: four point amplitudes}
\tl\cA_{J_1\dots J_4 }^{\De_1,\dots,\De_4}(0,\infty,1,z_4)=
\frac{\pi}{2}\mathcal{U}(\b_i)\delta\left(\zb_4-z_4\right)\delta\left(i(\ka-4)+\sum\limits_{i=1}^4\lambda_i\right)
\left(\frac{z_4-1}{-z_4 \left(\epsilon _{1,3}-1\right)+\epsilon _{1,4}-1}\right)^{i\lambda_1}
\\\x
\left(-\frac{z_4 \left(z_4-2\right)+1}{\left(z_4-1\right) \epsilon
	_{1,2}-z_4 \epsilon _{2,3}+\epsilon _{2,4}}\right)^{i\lambda_2} 
\left(\frac{z_4}{z_4 \left(-\epsilon _{1,3}\right)+\epsilon _{1,3}-\epsilon
	_{3,4}+z_4}\right)^{i\lambda_3}
\left(\frac{1}{\left(z_4-1\right) \epsilon _{1,4}-z_4 \epsilon _{3,4}+1}\right)^{i\lambda_4} 
\\\x 
\lim\limits_{L\rightarrow\infty}
\Amp{}{j_1\dots j_4}{\frac{z_4-1}{-z_4 \left(\epsilon _{1,3}-1\right)+\epsilon _{1,4}-1},\;-L^{-2}\frac{z_4 \left(z_4-2\right)+1}{\left(z_4-1\right) \epsilon
		_{1,2}-z_4 \epsilon _{2,3}+\epsilon _{2,4}},\;\frac{z_4}{z_4 \left(-\epsilon _{1,3}\right)+\epsilon _{1,3}-\epsilon
		_{3,4}+z_4},\;\frac{1}{\left(z_4-1\right) \epsilon _{1,4}-z_4 \epsilon _{3,4}+1}}{0,\;L,\;1,\; z_4}
\end{multline}
where we defined the shorthand notation
\be 
\e_{i_1,i_2\dots i_n}\equiv \e_{i_1}\e_{i_2}\cdots \e_{i_n}
\ee 
and where we have used the prescription detailed in Appendix~\ref{sec: generalized Cramer's rule} to compute $\b_i$. With $\b_i$, we can also compute $\mathcal{U}(\b_i)$ explicitly as can be seen in Table~\ref{eq: support function for 4 point amplitudes}.

The computation of the last term in \equref{eq: four point amplitudes} is straightforward, but we can simplify it even further with the following prescription. Given \emph{any} four point amplitude  of the form
\be 
\cA_{j_1\dots j_4}(p_i)=\sum\limits_{1\le i< j\le 4}\<ij\>^{m_{ij}}[ij]^{n_{ij}}\delta^4\left(\sum\limits_{i=1}^np_i^\mu\right)
\ee 
we can immediately write 
\be 
\lim\limits_{L\rightarrow\infty}
\Amp{}{j_1\dots j_4}{\frac{z_4-1}{-z_4 \left(\epsilon _{1,3}-1\right)+\epsilon _{1,4}-1},\;-L^{-2}\frac{z_4 \left(z_4-2\right)+1}{\left(z_4-1\right) \epsilon
		_{1,2}-z_4 \epsilon _{2,3}+\epsilon _{2,4}},\;\frac{z_4}{z_4 \left(-\epsilon _{1,3}\right)+\epsilon _{1,3}-\epsilon
		_{3,4}+z_4},\;\frac{1}{\left(z_4-1\right) \epsilon _{1,4}-z_4 \epsilon _{3,4}+1}}{0,\;L,\;1,\; z_4}
=
\sum\limits_{1\le i< j\le 4}
(-\e_{i,j})^{m_{ij}}\left(2a_{ij}^{(4)}\right)^{m_{ij}+n_{ij}}
\ee 
for 
\be 
\label{eq: four point amplitude prescription}
a_{12}^{(4)}=&-\sqrt{-\frac{\left(z_4-1\right){}^3 \epsilon _{1,2}}{\left(\left(z_4-1\right) \epsilon _{4,1,3}-z_4 \epsilon
		_4+\epsilon _3\right)^2}}
&
a_{13}^{(4)}=&-\sqrt{-\frac{\left(z_4-1\right) z_4 \epsilon _{1,3}}{\left(\left(z_4-1\right) \epsilon _{4,1,3}-z_4 \epsilon
		_4+\epsilon _3\right)^2}}
\\
a_{23}^{(4)}=&\sqrt{\frac{\left(z_4-1\right){}^2 z_4 \epsilon _{2,3}}{\left(\left(z_4-1\right) \epsilon _{4,1,3}-z_4 \epsilon
		_4+\epsilon _3\right)^2}}
&
a_{14}^{(4)}=&-z_4\sqrt{\frac{\left(z_4-1\right) \epsilon _{1,4}}{\left(\left(z_4-1\right) \epsilon _{4,1,3}-z_4 \epsilon
		_4+\epsilon _3\right)^2}}
\\
a_{24}^{(4)}=&\sqrt{-\frac{\left(z_4-1\right){}^2 \epsilon _{2,4}}{\left(\left(z_4-1\right) \epsilon _{4,1,3}-z_4 \epsilon
		_4+\epsilon _3\right)^2}}
&
a_{34}^{(4)}=&(1-z_4)\sqrt{-\frac{z_4 \epsilon _{3,4}}{\left(\left(z_4-1\right) \epsilon _{4,1,3}-z_4 \epsilon _4+\epsilon
		_3\right)^2}}
\ee 

This prescription is valid for any amplitude as it simply follows from the kinematics! For example, 
the color ordered gluon four point one loop amplitudes for all-plus and single-minus helicities in pure Yang-Mills theory
can be written as 
\be 
\cA^{\texttt{gluon}}_{++++}(p_i)=
-c\frac{[23][41]}{\<23\>\<41\>}
\delta^4\left(\sum\limits_{i=1}^np_i^\mu\right)\;,\quad
\cA^{\texttt{gluon}}_{-+++}(p_i)=
c\frac{\<24\>[24]^3}{[12]\<23\>\<34\>[41]}
\delta^4\left(\sum\limits_{i=1}^np_i^\mu\right)
\ee 
as seen in  \cite{Bern:2005hs,Mahlon:1993fe, Bern:1993qk}.\footnote{\label{footnote for c}Here, the coefficient $c=i\frac{N_p}{96\pi^2}$ where $N_p$ is the net number of states circulating in the loop.}  With the prescription above, we get
\footnotesize
\bea[eq: 4point loop]
\lim\limits_{L\rightarrow\infty}
\Amp{\texttt{gluon}}{++++}{\frac{z_4-1}{-z_4 \left(\epsilon _{1,3}-1\right)+\epsilon _{1,4}-1},\;-L^{-2}\frac{z_4 \left(z_4-2\right)+1}{\left(z_4-1\right) \epsilon
		_{1,2}-z_4 \epsilon _{2,3}+\epsilon _{2,4}},\;\frac{z_4}{z_4 \left(-\epsilon _{1,3}\right)+\epsilon _{1,3}-\epsilon
		_{3,4}+z_4},\;\frac{1}{\left(z_4-1\right) \epsilon _{1,4}-z_4 \epsilon _{3,4}+1}}{0,\;L,\;1,\; z_4}
=&-\frac{c}{\e_{1,2,3,4}}
\\
\lim\limits_{L\rightarrow\infty}
\Amp{\texttt{gluon}}{-+++}{\frac{z_4-1}{-z_4 \left(\epsilon _{1,3}-1\right)+\epsilon _{1,4}-1},\;-L^{-2}\frac{z_4 \left(z_4-2\right)+1}{\left(z_4-1\right) \epsilon
		_{1,2}-z_4 \epsilon _{2,3}+\epsilon _{2,4}},\;\frac{z_4}{z_4 \left(-\epsilon _{1,3}\right)+\epsilon _{1,3}-\epsilon
		_{3,4}+z_4},\;\frac{1}{\left(z_4-1\right) \epsilon _{1,4}-z_4 \epsilon _{3,4}+1}}{0,\;L,\;1,\; z_4}
=&\frac{c\;\mathtt{sgn}\left(z_4(1-z_4)\right)}{z_4^2}
\eea
\normalsize
We can also consider one loop gravity amplitudes for both all-positive and one-minus cases. These amplitudes have been computed using string based methods \cite{Bern:1993sx,  Bern:1993wt, Dunbar:1994bn}. Also, it is interesting to note that the four-point all-plus one-loop gravity amplitude can be calculated using the BCJ double copy construction and we refer the readers to \cite{Boels:2013bi} for more details. In spinor helicity formalism, they can be written in a compact form \cite{Boels:2013bi,Bern:1998sv}:\footnote{The parameters $s$, $t$, and $u$ are the standard Mandelstam variables, e.g. $s=\<12\>[12]$.}
\be 
\cA^{\texttt{graviton}}_{++++}(p_i)=&
-\frac{i}{(4\pi)^2}\frac{s^2+t^2+u^2}{120}\left(\frac{s t}{\<12\>\<23\>\<34\>\<41\>}\right)^2
\delta^4\left(\sum\limits_{i=1}^np_i^\mu\right)\\
\cA^{\texttt{graviton}}_{-+++}(p_i)=&4\left(\frac{st}{u}\right)^2\left(\frac{s^2+st+t^2}{5760}\right)^2\left(\frac{[24]^2}{[12]\<23\>\<34\>[41]}\right)^2\delta^4\left(\sum\limits_{i=1}^np_i^\mu\right)
\ee
With the prescription above, we get
\footnotesize
\bea[eq: 4point loop 2]
\lim\limits_{L\rightarrow\infty}
\Amp{\texttt{graviton}}{++++}{\frac{z_4-1}{-z_4 \left(\epsilon _{1,3}-1\right)+\epsilon _{1,4}-1},\;-L^{-2}\frac{z_4 \left(z_4-2\right)+1}{\left(z_4-1\right) \epsilon
		_{1,2}-z_4 \epsilon _{2,3}+\epsilon _{2,4}},\;\frac{z_4}{z_4 \left(-\epsilon _{1,3}\right)+\epsilon _{1,3}-\epsilon
		_{3,4}+z_4},\;\frac{1}{\left(z_4-1\right) \epsilon _{1,4}-z_4 \epsilon _{3,4}+1}}{0,\;L,\;1,\; z_4}
\nn\\
=-\frac{2i}{15(4\pi)^2}\frac{ \left(z_4-1\right){}^2 \left(z_4 \left(z_4 \left(2 \left(z_4-2\right) z_4+7\right)-4\right)+1\right)}{z_4^2 \left(\left(z_4-1\right)
	\epsilon _{1,3,4}-z_4 \epsilon _4+\epsilon _3\right)^4}
\\
\lim\limits_{L\rightarrow\infty}
\Amp{\texttt{graviton}}{-+++}{\frac{z_4-1}{-z_4 \left(\epsilon _{1,3}-1\right)+\epsilon _{1,4}-1},\;-L^{-2}\frac{z_4 \left(z_4-2\right)+1}{\left(z_4-1\right) \epsilon
		_{1,2}-z_4 \epsilon _{2,3}+\epsilon _{2,4}},\;\frac{z_4}{z_4 \left(-\epsilon _{1,3}\right)+\epsilon _{1,3}-\epsilon
		_{3,4}+z_4},\;\frac{1}{\left(z_4-1\right) \epsilon _{1,4}-z_4 \epsilon _{3,4}+1}}{0,\;L,\;1,\; z_4}
\nn\\=\frac{25}{60^4}\frac{\left(z_4-1\right){}^4 \left(\left(z_4-1\right){}^2 \epsilon _{2,3} \left| z_4\right| +z_4 \left(\left(z_4-3\right) z_4+4\right)
	\left(z_4-1\right)+1\right)}{z_4^6 \left(\left(z_4-1\right) \epsilon _{1,3,4}-z_4 \epsilon _4+\epsilon _3\right)^4}
\eea
\normalsize

We obtain the full celestial amplitudes by inserting \equref{eq: 4point loop} and \equref{eq: 4point loop 2} into \equref{eq: four point amplitudes}. For example,
\begin{multline}
\left(\tl\cA^{\texttt{gluon}}\right)_{++++ }^{\De_1,\dots,\De_4}(0,\infty,1,z_4)=
-\frac{\pi c}{2 \e_{1,2,3,4}}\mathcal{U}(\b_i)\delta\left(\zb_4-z_4\right)\delta\left(\sum\limits_{i=1}^4\lambda_i\right)
\\\x
\left(\frac{z_4-1}{-z_4 \left(\epsilon _{1,3}-1\right)+\epsilon _{1,4}-1}\right)^{i\lambda_1}
\left(-\frac{z_4 \left(z_4-2\right)+1}{\left(z_4-1\right) \epsilon
	_{1,2}-z_4 \epsilon _{2,3}+\epsilon _{2,4}}\right)^{i\lambda_2} 
\\\x
\left(\frac{z_4}{z_4 \left(-\epsilon _{1,3}\right)+\epsilon _{1,3}-\epsilon
	_{3,4}+z_4}\right)^{i\lambda_3}
\left(\frac{1}{\left(z_4-1\right) \epsilon _{1,4}-z_4 \epsilon _{3,4}+1}\right)^{i\lambda_4} 
\end{multline}
For specific choices of incoming/outgoing momenta, the expression simplifies significantly; i.e.
\footnotesize
\be 
\left(\tl\cA^{\texttt{gluon}}\right)_{++++}^{\De_1,\dots,\De_4}(0,\infty,1,z_4)\evaluated_{\substack{p_{2,3}\text{ : incoming}\\p_{1,4}\text{ : outgoing}}}=\left\{\begin{aligned} &
	-\frac{\pi c}{2}
	\left(z_4-1\right){}^{i \left(\lambda _1+2 \lambda _2\right)} z_4^{i \lambda
		_3} 
	\delta \left(\zb_4-z_4\right)
	\delta\left(\sum\limits_{i=1}^4\lambda_i\right) \qquad& z_4\ge 1\\ &
	-\frac{\pi c}{2}
	\left(z_4-1\right){}^{i \left(\lambda _1+2 \lambda _2\right)} z_4^{i \lambda
		_3} 
	\delta \left(\zb_4-z_4\right)
	\delta\left(\sum\limits_{i=1}^4\lambda_i\right)e^{2\pi\lambda_2} \qquad& 1>z_4\ge\half
\end{aligned}\right.
\ee 
\normalsize
We would like to remind the reader that one can get the standard form of the amplitude, i.e. $\left(\tl\cA^{\texttt{gluon}}\right)_{++++}^{\De_1,\dots,\De_4}(\chi_1,\chi_2,\chi_3,\chi_4)$, from its form in the conformal frame using  \equref{eq: pulling back from conformal frame}.

\subsection{Five point amplitudes}
\label{sec: five point loop}

\begin{table}
	\centering
	\caption{\label{eq: support function for 5 point amplitudes} The break-down of the support of the five point celestial amplitude on $\left\{(x_4,y_4,x_5,y_z)\in\R^4\;|\;z_4=x_4+iy_4,z_5=x_5+iy_5\right\}$ depending on which momenta lie on future lightcone ($\e=1$) and which momenta lie on past lightcone ($\e=-1$).}
	\footnotesize
	\begin{tabular}{cc}
		\hline\hline
		\textbf{Case} & \textbf{Region }$\mathcal{U}(\b_i)$\textbf{ is }$1$
		\\\hline\\[-.12in]
		$\e_1=\e_3=\e_4=-\e_5$& $x_4<\frac{x_5 y_4}{y_5}\land \frac{\left(x_5-2\right) y_4}{y_5}<x_4\land \left(\left(2
		y_4<y_5\land y_5<0\right)\lor \left(2 y_4>y_5\land y_5>0\right)\right)$ \\[.05in]\hline\\[-.12in]
		$\e_1=\e_3=-\e_4=\e_5$& $\frac{x_5 y_4}{y_5}<x_4\land x_4<\frac{x_5 y_4}{y_5}+2\land \left(\left(y_4>0\land y_4<2
		y_5\right)\lor \left(y_4<0\land 2 y_5<y_4\right)\right) $\\[.05in]\hline\\[-.12in]
		$\e_1=\e_3=-\e_4=-\e_5$& $x_4<\frac{\left(x_5-2\right) y_4}{y_5}+2\land \frac{x_5 y_4}{y_5}<x_4\land
		\left(\left(y_5>0\land y_4<0\right)\lor \left(y_4>0\land y_5<0\right)\right) $\\[.05in]\hline\\[-.12in]
		$\e_1=-\e_3=\e_4=\e_5$& $\begin{aligned}
		\left(x_4>\frac{x_5 y_4}{y_5}+\frac{1}{2}\land \left(\left(y_4>0\land y_4+y_5\leq 0\right)\lor
		\left(y_4<0\land y_4+y_5\geq 0\right)\right)\right)\\\lor \left(y_5 \left(-2 x_5 y_4+2 x_4 y_5+y_4\right)>0\land
		\left(\left(y_4+y_5>0\land y_5<0\right)\lor \left(y_5>0\land y_4+y_5<0\right)\right)\right)
		\end{aligned}$ \\[.05in]\hline\\[-.12in]
		$\e_1=-\e_3=\e_4=-\e_5$& $\begin{aligned}
		\left(\left(y_4<0\lor 2 y_4\leq y_5\right)\land x_4>\frac{\left(x_5-1\right)
			y_4}{y_5}+\frac{1}{2}\land \left(y_4>0\lor 2 y_4>y_5\right)\right)\\\lor \left(\left(y_5<0\lor 2
		y_4>y_5\right)\land x_4>\frac{x_5 y_4}{y_5}\land \left(2 y_4\leq y_5\lor y_5>0\right)\right) 
		\end{aligned}$\\[.05in]\hline\\[-.12in]
		$\e_1=-\e_3=-\e_4=\e_5$& $\begin{aligned}
		\left(x_4<\frac{x_5 y_4}{y_5}\land \left(\left(y_4>0\land y_4<2 y_5\right)\lor \left(y_4<0\land
		2 y_5<y_4\right)\right)\right)\\\lor \left(x_4<\frac{\left(2 x_5-1\right) y_4}{2 y_5}+1\land
		\left(\left(y_5>0\land y_4\geq 2 y_5\right)\lor \left(y_5<0\land y_4\leq 2 y_5\right)\right)\right) 
		\end{aligned}$\\[.05in]\hline\\[-.12in]
		$\e_1=-\e_3=-\e_4=-\e_5 $& $x_4<\frac{x_5 y_4}{y_5}\land \left(\left(y_5>0\land y_4<0\right)\lor \left(y_4>0\land
		y_5<0\right)\right) $
		\\[.05in]\hline\hline
	\end{tabular}
\end{table}
After considering four point, we would like to extend the computation of celestial amplitudes to five points. Again, we will focus our attention to all plus and single minus results. We have seen that a generic five point  celestial amplitude can be computed from \equref{eq: celestial amplitudes final form 2}, which becomes
\small
\begin{multline}
\label{eq: five point amplitudes}
\tl\cA_{J_1\dots J_5 }^{\De_1,\dots,\De_5}(0,\infty,1,z_4,z_5)=
\frac{\pi}{2}\frac{\mathcal{U}(\b_i)}{\abs{\varphi}}\delta\left(i(\kappa-5)+\sum\limits_{i=1}^5\lambda_i\right)
\left(\frac{\left(-z_5 \bar{z}_4+\bar{z}_4+\left(z_4-1\right) \bar{z}_5-z_4+z_5\right) \epsilon _{1,3,4,5}}{\varphi}\right)^{i\lambda_1}
\\\x 
\left(\frac{\left(\bar{z}_4+z_5 \left(\bar{z}_4 \left(\bar{z}_5-2\right)+1\right)-\bar{z}_5+z_4 \left(z_5
	\bar{z}_4-\left(\bar{z}_4+z_5-2\right) \bar{z}_5-1\right)\right) \epsilon _{2,3,4,5}}{\varphi}\right)^{i\lambda_2}
\left(\frac{\left(z_5 \bar{z}_4-z_4 \bar{z}_5\right) \epsilon _{4,5}}{\varphi}\right)^{i\lambda_3}
\\\x
\left(\frac{\left(\bar{z}_5-z_5\right) \epsilon _{3,5}}{\varphi}\right)^{i\lambda_4}\left(\frac{\left(z_4-\bar{z}_4\right) \epsilon _{3,4}}{\varphi}\right)^{i\lambda_5}
\lim\limits_{L\rightarrow\infty} 
\Amp{}{j_1\dots j_n}{1-\sum\limits_{i=1}^3\b_i,\;-L^{-2}\sum\limits_{i=1}^3\epsilon_2\epsilon_{i+2} \b_{i}\left(1+z_{i+2}\zb_{i+2}\right),\;\b_1,\;\b_2,\;\b_3}{0,\;L,\;1,\; z_4,\;z_5}
\end{multline}
\normalsize
for
\be 
\label{eq: definition of varphi}
\varphi \equiv \left(\bar{z}_5-z_5\right) \left(\epsilon _{3,5}-\epsilon _{1,3,4,5}\right)+\bar{z}_4 \left(z_5 \epsilon
_{4,5}-\left(z_5-1\right) \epsilon _{1,3,4,5}-\epsilon _{3,4}\right)+z_4 \left(-\bar{z}_5 \epsilon
_{4,5}+\left(\bar{z}_5-1\right) \epsilon _{1,3,4,5}+\epsilon _{3,4}\right)
\ee 
where we have used the prescription detailed in Appendix~\ref{sec: generalized Cramer's rule} to compute $\b_i$. With $\b_i$, we can also compute $\mathcal{U}(\b_i)$ explicitly as can be seen in Table~\ref{eq: support function for 5 point amplitudes}.

We can provide a prescription to compute the last term in \equref{eq: five point amplitudes}, similar to what we did for four point amplitudes in \equref{eq: four point amplitude prescription}. Given \emph{any} five point amplitude  of the form
\be 
\cA_{j_1\dots j_5}(p_i)=\sum\limits_{1\le i< j\le 5}\<ij\>^{m_{ij}}[ij]^{n_{ij}}\delta^4\left(\sum\limits_{i=1}^np_i^\mu\right)
\ee 
we have 
\be 
\label{eq: five point prescription}
\lim\limits_{L\rightarrow\infty} 
\Amp{}{j_1\dots j_n}{1-\sum\limits_{i=1}^3\b_i,\;-L^{-2}\sum\limits_{i=1}^3\epsilon_2\epsilon_{i+2} \b_{i}\left(1+z_{i+2}\zb_{i+2}\right),\;\b_1,\;\b_2,\;\b_3}{0,\;L,\;1,\; z_4,\;z_5}
=
\sum\limits_{1\le i< j\le 5}
\left(-\frac{\e_{i,j}z_{ij}}{\zb_{ij}}\right)^{m_{ij}}\left(2a^{(5)}_{ij}\right)^{m_{ij}+n_{ij}}
\ee 
for  the coefficients $a_{ij}^{(5)}$ given in \equref{eq: five point amplitude prescription}.

By inserting $\lim\limits_{L\rightarrow\infty} 
\Amp{}{j_1\dots j_n}{\cdots}{\cdots}$ into \equref{eq: five point amplitudes}, we can obtain the celestial form of \emph{any} amplitude. As an example, we know that the color ordered gluon five point amplitude in pure Yang-Mills theory reads in spinor helicity variables as 
\be
\label{eq: five point amplitude}
\cA^{\texttt{gluon}}_{+++++}(p_i)=&-c\frac{\sum\limits_{1\le i< j< k<l\le 5}\<ij\>[jk]\<kl\>[li]}{\<12\>\<23\>\<34\>\<45\>\<51\>}\delta^4\left(\sum\limits_{i=1}^np_i^\mu\right)
\\
\cA^{\texttt{gluon}}_{-++++}(p_i)=&
\frac{c}{\<34\>^2}
\left(
-\frac{[25]^3}{[12][51]}+\frac{\<14\>^3[45]\<35\>}{\<12\>\<23\>\<45\>^2}-\frac{\<13\>^3[32]\<42\>}{\<15\>\<54\>\<32\>^2}
\right)
\delta^4\left(\sum\limits_{i=1}^np_i^\mu\right)
\ee
where $c$ is given in footnote~\ref{footnote for c}. With the prescription above, we can find the full expression as given in \equref{eq: five point result}.

For special configurations, the expressions simplify; for example,
\small
\begin{multline}
\lim\limits_{L\rightarrow\infty} 
\Amp{\texttt{gluon}}{-++++}{1-\sum\limits_{i=1}^3\b_i,\;-L^{-2}\sum\limits_{i=1}^3\epsilon_2\epsilon_{i+2} \b_{i}\left(1+z_{i+2}\zb_{i+2}\right),\;\b_1,\;\b_2,\;\b_3}{0,\;L,\;1,\; z_4,\;z_5}_{\substack{p_{1,2}\text{ : incoming}\\p_{3,4,5}\text{ : outgoing}}}
\\=
\frac{2 \left| \frac{z_5 \bar{z}_4-z_4 \bar{z}_5}{2 z_4-2 \bar{z}_4}\right| }{z_5
	\left(z_5-z_4\right)}+\frac{\left| \frac{\left(z_4-\bar{z}_4\right) \left(\bar{z}_4 \left(\bar{z}_5-2\right) z_5+z_5+\bar{z}_4-\bar{z}_5+z_4
		\left(z_5 \left(\bar{z}_4-\bar{z}_5\right)-\left(\bar{z}_4-2\right) \bar{z}_5-1\right)\right)}{\left(-\bar{z}_4 z_5+z_5+\bar{z}_4+z_4
		\left(\bar{z}_5-1\right)-\bar{z}_5\right){}^2}\right| }{\bar{z}_5}
	\\-\frac{\left(z_5-1\right) z_4^3 \left(\bar{z}_4-\bar{z}_5\right) \left| \frac{\bar{z}_5-z_5}{\bar{z}_4 \left(\bar{z}_5-2\right)
			z_5+z_5+\bar{z}_4-\bar{z}_5+z_4 \left(z_5 \left(\bar{z}_4-\bar{z}_5\right)-\left(\bar{z}_4-2\right) \bar{z}_5-1\right)}\right|
	}{\left(z_4-z_5\right)^2}
\end{multline}
\normalsize
with which the full result in \equref{eq: five point amplitudes} becomes
\begin{multline}
\left(\tl\cA^{\texttt{gluon}}\right)_{-++++}^{\De_1,\dots,\De_5}(0,\infty,1,z_4,z_5)\evaluated_{{\substack{p_{1,2}\text{ : incoming}\\p_{3,4,5}\text{ : outgoing}}}}=
\frac{\pi}{4}\frac{\mathcal{U}(\b_i)}{\left| z_5 \left(\bar{z}_4-1\right)-\bar{z}_4-z_4 \left(\bar{z}_5-1\right)+\bar{z}_5\right|}\delta\left(\sum\limits_{i=1}^5\lambda_i\right)
\\\x 
\left(\frac{\bar{z}_4+z_5 \left(\bar{z}_4 \left(\bar{z}_5-2\right)+1\right)-\bar{z}_5+z_4 \left(z_5 \bar{z}_4-\left(\bar{z}_4+z_5-2\right)
	\bar{z}_5-1\right)}{-z_5 \bar{z}_4+\bar{z}_4+\left(z_4-1\right) \bar{z}_5-z_4+z_5}\right)^{i\lambda_2}
\left(\frac{z_4 \bar{z}_5-z_5 \bar{z}_4}{-z_5 \bar{z}_4+\bar{z}_4+\left(z_4-1\right) \bar{z}_5-z_4+z_5}\right)^{i\lambda_3}
\\\x
\left(\frac{z_5-\bar{z}_5}{-z_5 \bar{z}_4+\bar{z}_4+\left(z_4-1\right) \bar{z}_5-z_4+z_5}\right)^{i\lambda_4}\left(\frac{\bar{z}_4-z_4}{-z_5 \bar{z}_4+\bar{z}_4+\left(z_4-1\right) \bar{z}_5-z_4+z_5}\right)^{i\lambda_5}
\\\x\Bigg(
\frac{2 \left| \frac{z_5 \bar{z}_4-z_4 \bar{z}_5}{2 z_4-2 \bar{z}_4}\right| }{z_5
	\left(z_5-z_4\right)}+\frac{\left| \frac{\left(z_4-\bar{z}_4\right) \left(\bar{z}_4 \left(\bar{z}_5-2\right) z_5+z_5+\bar{z}_4-\bar{z}_5+z_4
		\left(z_5 \left(\bar{z}_4-\bar{z}_5\right)-\left(\bar{z}_4-2\right) \bar{z}_5-1\right)\right)}{\left(-\bar{z}_4 z_5+z_5+\bar{z}_4+z_4
		\left(\bar{z}_5-1\right)-\bar{z}_5\right){}^2}\right| }{\bar{z}_5}
\\-\frac{\left(z_5-1\right) z_4^3 \left(\bar{z}_4-\bar{z}_5\right) \left| \frac{\bar{z}_5-z_5}{\bar{z}_4 \left(\bar{z}_5-2\right)
		z_5+z_5+\bar{z}_4-\bar{z}_5+z_4 \left(z_5 \left(\bar{z}_4-\bar{z}_5\right)-\left(\bar{z}_4-2\right) \bar{z}_5-1\right)}\right|
}{\left(z_4-z_5\right)^2}
\Bigg)
\end{multline}

As another explicit example, we can consider a five point graviton amplitude. For the all-plus rational loop amplitude, we have
\begin{multline}
\label{eq: five point graviton amplitude}
\cA^{\texttt{graviton}}_{+++++}(p_i)=\frac{i}{960(4\pi)^2}
\Bigg(
\\
\frac{[45]\left(
\<12\>[12]\<34\>[34]-\left(2\leftrightarrow 3\right)+\left(2\leftrightarrow 4\right)
+\left(4\leftrightarrow 5\right)-\left(\begin{aligned}
4\leftrightarrow 5\\2\leftrightarrow 3
\end{aligned}\right)+\left(\begin{aligned}
2\rightarrow 5\\4\rightarrow 2
\end{aligned}\right)
\right)^3}{\<12\>^2\<23\>^2\<45\>\<34\>\<41\>\<35\>\<51\>}
\\+\text{permutations}
\Bigg)\delta^4\left(\sum\limits_{i=1}^np_i^\mu\right)
\end{multline}
where there are 30 distinct permutation in total. Below, we will consider this term alone; one can analogously repeat the computation for permuted terms; for details of the derivation of \equref{eq: five point graviton amplitude} and for further information on the other permutations, see eqn.~4.23 of \cite{Bern:1998sv}.

For generic $\epsilon_i$, the expression $\lim\limits_{L\rightarrow\infty} 
\Amp{\texttt{graviton}}{+++++}{\cdots}{\cdots}$ takes a rather complicated form; however, it becomes manageable if we switch to real parameters $\{(x_k,y_k)\in\R^2\;|\;z_i=x_k+iy_k\}$:
\scriptsize
\begin{multline}
\lim\limits_{L\rightarrow\infty} 
\Amp{\texttt{graviton}}{-++++}{1-\sum\limits_{i=1}^3\b_i,\;-L^{-2}\sum\limits_{i=1}^3\epsilon_2\epsilon_{i+2} \b_{i}\left(1+x_{i+2}^2+y^2_{i+2}\right),\;\b_1,\;\b_2,\;\b_3}{0,\;L,\;1,\; x_4+iy_4,\;x_5+iy_5}
\\=
-\frac{128 \epsilon _{4,5} \left(x_4-x_5-i \left(y_4-y_5\right)\right) \left(y_5 \left(x_4 \left(\epsilon _{4,5}-\epsilon
	_{1,3,4,5}\right)-\epsilon _{3,5}+\epsilon _{1,3,4,5}\right)+y_4 \left(x_5 \left(\epsilon _{1,3,4,5}-\epsilon _{4,5}\right)+\epsilon
	_{3,4}-\epsilon _{1,3,4,5}\right)\right)^8 }{\left| y_4 y_5\right|  \left(x_4+i
	y_4-1\right) \left(x_4+i y_4\right) \left(x_4-x_5+i \left(y_4-y_5\right)\right) \left(x_5+i y_5-1\right) \left(x_5+i y_5\right)
	\left(\left(x_5-1\right) y_4-\left(x_4-1\right) y_5\right)^2 \left(x_5 y_4-x_4 y_5\right)^2}
\\\x \frac{\left(\frac{\epsilon _{1,2,3,4} \left(x_4^2-x_4+y_4^2\right) \left| y_5 \left(\left(x_5-1\right)
		y_4-\left(x_4-1\right) y_5\right) \left(x_5 y_4-x_4 y_5\right) \left(y_5 \left(x_4-1\right){}^2+y_4^2 y_5-y_4 \left(x_5^2-2
		x_5+y_5^2+1\right)\right)\right| }{\left(y_5 \left(x_4 \left(\epsilon _{4,5}-\epsilon _{1,3,4,5}\right)-\epsilon _{3,5}+\epsilon
		_{1,3,4,5}\right)+y_4 \left(x_5 \left(\epsilon _{1,3,4,5}-\epsilon _{4,5}\right)+\epsilon _{3,4}-\epsilon
		_{1,3,4,5}\right)\right)^4}+\frac{\epsilon _{1,2,3,5} \left(x_5^2-x_5+y_5^2\right) \left| y_4 \left(\left(x_5-1\right) y_4-\left(x_4-1\right)
		y_5\right) \left(x_5 y_4-x_4 y_5\right) \left(y_5 \left(x_4-1\right){}^2+y_4^2 y_5-y_4 \left(x_5^2-2 x_5+y_5^2+1\right)\right)\right|
	}{\left(y_5 \left(x_4 \left(\epsilon _{4,5}-\epsilon _{1,3,4,5}\right)-\epsilon _{3,5}+\epsilon _{1,3,4,5}\right)+y_4 \left(x_5 \left(\epsilon
		_{1,3,4,5}-\epsilon _{4,5}\right)+\epsilon _{3,4}-\epsilon _{1,3,4,5}\right)\right)^4}\right)^3}{\left(\left(x_4-1\right)^2 y_5-y_4
	\left(x_5^2-2 x_5+y_5^2+1\right)+y_4^2 y_5\right)^2}
\\+\text{contributions due to permuted terms}
\end{multline}
\normalsize
If we consider a specific configuration of incoming/outgoing momenta, the expression simplifies enough so that we can write it again in terms of $z_i$ and $\zb_i$; for instance,
\scriptsize
\begin{multline}
\lim\limits_{L\rightarrow\infty} 
\Amp{\texttt{graviton}}{-++++}{1-\sum\limits_{i=1}^3\b_i,\;-L^{-2}\sum\limits_{i=1}^3\epsilon_2\epsilon_{i+2} \b_{i}\left(1+z_{i+2}\zb_{i+2}\right),\;\b_1,\;\b_2,\;\b_3}{0,\;L,\;1,\; z_4,\;z_5}_{\substack{p_{1,2}\text{ : incoming}\\p_{3,4,5}\text{ : outgoing}}}
\\=-
\frac{\left(\bar{z}_4-\bar{z}_5\right) \left| \frac{1}{\left(z_4-\bar{z}_4\right) \left(z_5-\bar{z}_5\right)}\right|}{\left(z_4-1\right) z_4
	\left(z_4-z_5\right) \left(z_5-1\right) z_5 \left(z_5 \left(-\bar{z}_4\right)+\bar{z}_4+z_4 \left(\bar{z}_5-1\right)-\bar{z}_5+z_5\right){}^3
	\left(z_5 \bar{z}_4-z_4 \bar{z}_5\right)^2 \left(z_5 \bar{z}_4 \left(\bar{z}_5-2\right)+\bar{z}_4-\bar{z}_5+z_4 \left(z_5
	\left(\bar{z}_4-\bar{z}_5\right)-\left(\bar{z}_4-2\right) \bar{z}_5-1\right)+z_5\right)^2}
\\\x 
\Bigg(\left(z_4 \left(2
\bar{z}_4-1\right)-\bar{z}_4\right) \left| \left(z_5-\bar{z}_5\right) \left(z_5 \bar{z}_4-z_4 \bar{z}_5\right) \left(\bar{z}_4
\left(\bar{z}_5-2\right) z_5+z_5+\bar{z}_4-\bar{z}_5+z_4 \left(z_5 \left(\bar{z}_4-\bar{z}_5\right)-\left(\bar{z}_4-2\right)
\bar{z}_5-1\right)\right)\right|
\\
+\left(z_5 \left(2 \bar{z}_5-1\right)-\bar{z}_5\right) \left| \left(z_4-\bar{z}_4\right) \left(z_5
\bar{z}_4-z_4 \bar{z}_5\right) \left(\bar{z}_4 \left(\bar{z}_5-2\right) z_5+z_5+\bar{z}_4-\bar{z}_5+z_4 \left(z_5
\left(\bar{z}_4-\bar{z}_5\right)-\left(\bar{z}_4-2\right) \bar{z}_5-1\right)\right)\right| \Bigg)^3
\\+\text{contributions due to permuted terms}
\end{multline}
\normalsize
with which the celestial amplitude becomes
\scriptsize
\begin{multline}
\label{eq: 5 point graviton}
\left(\tl\cA^{\texttt{graviton}}\right)_{-++++}^{\De_1,\dots,\De_5}(0,\infty,1,z_4,z_5)\evaluated_{{\substack{p_{1,2}\text{ : incoming}\\p_{3,4,5}\text{ : outgoing}}}}=
-\frac{\pi}{4}\frac{\mathcal{U}(\b_i)}{\left| z_5 \left(\bar{z}_4-1\right)-\bar{z}_4-z_4 \left(\bar{z}_5-1\right)+\bar{z}_5\right|}\delta\left(\sum\limits_{i=1}^5\lambda_i\right)\left(\frac{z_5-\bar{z}_5}{-z_5 \bar{z}_4+\bar{z}_4+\left(z_4-1\right) \bar{z}_5-z_4+z_5}\right)^{i\lambda_4}
\\\x 
\left(\frac{\bar{z}_4+z_5 \left(\bar{z}_4 \left(\bar{z}_5-2\right)+1\right)-\bar{z}_5+z_4 \left(z_5 \bar{z}_4-\left(\bar{z}_4+z_5-2\right)
	\bar{z}_5-1\right)}{-z_5 \bar{z}_4+\bar{z}_4+\left(z_4-1\right) \bar{z}_5-z_4+z_5}\right)^{i\lambda_2}
\left(\frac{z_4 \bar{z}_5-z_5 \bar{z}_4}{-z_5 \bar{z}_4+\bar{z}_4+\left(z_4-1\right) \bar{z}_5-z_4+z_5}\right)^{i\lambda_3}
\left(\frac{\bar{z}_4-z_4}{-z_5 \bar{z}_4+\bar{z}_4+\left(z_4-1\right) \bar{z}_5-z_4+z_5}\right)^{i\lambda_5}
\\\x 
\frac{\left(\bar{z}_4-\bar{z}_5\right) \left| \frac{1}{\left(z_4-\bar{z}_4\right) \left(z_5-\bar{z}_5\right)}\right|}{\left(z_4-1\right) z_4
	\left(z_4-z_5\right) \left(z_5-1\right) z_5 \left(z_5 \left(-\bar{z}_4\right)+\bar{z}_4+z_4 \left(\bar{z}_5-1\right)-\bar{z}_5+z_5\right){}^3
	\left(z_5 \bar{z}_4-z_4 \bar{z}_5\right)^2 \left(z_5 \bar{z}_4 \left(\bar{z}_5-2\right)+\bar{z}_4-\bar{z}_5+z_4 \left(z_5
	\left(\bar{z}_4-\bar{z}_5\right)-\left(\bar{z}_4-2\right) \bar{z}_5-1\right)+z_5\right)^2}
\\\x 
\Bigg(\left(z_4 \left(2
\bar{z}_4-1\right)-\bar{z}_4\right) \left| \left(z_5-\bar{z}_5\right) \left(z_5 \bar{z}_4-z_4 \bar{z}_5\right) \left(\bar{z}_4
\left(\bar{z}_5-2\right) z_5+z_5+\bar{z}_4-\bar{z}_5+z_4 \left(z_5 \left(\bar{z}_4-\bar{z}_5\right)-\left(\bar{z}_4-2\right)
\bar{z}_5-1\right)\right)\right|
\\
+\left(z_5 \left(2 \bar{z}_5-1\right)-\bar{z}_5\right) \left| \left(z_4-\bar{z}_4\right) \left(z_5
\bar{z}_4-z_4 \bar{z}_5\right) \left(\bar{z}_4 \left(\bar{z}_5-2\right) z_5+z_5+\bar{z}_4-\bar{z}_5+z_4 \left(z_5
\left(\bar{z}_4-\bar{z}_5\right)-\left(\bar{z}_4-2\right) \bar{z}_5-1\right)\right)\right| \Bigg)^3
\\+\text{contributions due to permuted terms}
\end{multline}
\normalsize

With \equref{eq: 5 point graviton}, we have concluded our series of explicit gluon and graviton celestial amplitude results. As we can see in \equref{eq: celestial amplitudes final form 1}, there are complicated integrations that need to be carried out beyond five points hence it is not practical to provide the full explicit answers for higher point amplitudes. Nevertheless, in next section, we will discuss their generic forms and provide an explicit integrand for all plus one loop gluon amplitude.

\subsection{Higher point amplitudes}
\label{sec: higher point loop}

We have seen that a generic higher point  celestial amplitude can be computed from \equref{eq: celestial amplitudes final form 1}, which becomes
\footnotesize
\begin{multline}
\label{eq: higher point generic form}
\tl\cA_{J_1\dots J_n }^{\De_1,\dots,\De_n}(0,\infty,1,z_4,\dots,z_n)=
\frac{\pi}{2}\frac{\mathcal{U}(\b_i)}{\abs{\varphi}}\delta\left(i(\kappa-n)+\sum\limits_{i=1}^n\lambda_i\right)
\prod\limits_{k=6}^n\left(\int\limits_0^1d\sigma_k\sigma_k^{i\lambda_k}\right)\b_1^{i\lambda_3}\b_2^{i\lambda_4}\b_3^{i\lambda_5}
\\\x 
\left(1-\sum\limits_{i=1}^3\b_i-\sum\limits_{i=6}^n\s_i\right)^{i\lambda_1}
\left(-\sum\limits_{i=1}^3\epsilon_2\epsilon_{i+2} \b_i\left(1+z_{i+2}\zb_{i+2}\right)-\sum\limits_{i=6}^n\epsilon_2\epsilon_i \sigma_i\left(1+z_i\zb_i\right)\right)^{i\lambda_2}
\\\x 
\lim\limits_{L\rightarrow\infty} 
\Amp{}{j_1\dots j_n}{1-\sum\limits_{i=1}^3\b_i-\sum\limits_{i=6}^n\s_i,\;-L^{-2}\left(\sum\limits_{i=1}^3\epsilon_2\epsilon_{i+2} \b_i\left(1+z_{i+2}\zb_{i+2}\right)+\sum\limits_{i=6}^n\epsilon_2\epsilon_i \sigma_i\left(1+z_i\zb_i\right)\right),\;\b_1,\;\b_2,\;\b_3,\;\s_6,\;\dots,\;\s_n}{0,\;L,\;1,\; z_4,\;\dots,\; z_n}
\end{multline}
\normalsize
where $\varphi$ is defined in \equref{eq: definition of varphi}. One can also compute $\b_i$ straightforwardly as explained in Appendix~\ref{sec: generalized Cramer's rule}; for reader's convenience, we provide the explicit results: 
\bea
\varphi\b_1=&
\left(z_5 \bar{z}_4-z_4 \bar{z}_5\right) \epsilon _{4,5}+\sum _{i=6}^n \sigma _i \bigg(z_4 \left(\bar{z}_i \left(\epsilon _{4,i}-\epsilon _{1,4,5,i}\right)+\bar{z}_5 \left(\epsilon _{1,4,5,i}-\epsilon
_{4,5}\right)\right)\nn\\&+z_i \left(\bar{z}_5 \left(\epsilon _{5,i}-\epsilon _{1,4,5,i}\right)+\bar{z}_4 \left(\epsilon _{1,4,5,i}-\epsilon
_{4,i}\right)\right)+z_5 \left(\bar{z}_4 \left(\epsilon _{4,5}-\epsilon _{1,4,5,i}\right)+\bar{z}_i \left(\epsilon _{1,4,5,i}-\epsilon
_{5,i}\right)\right)\bigg)
\\
\varphi\b_2=&\left(\bar{z}_5-z_5\right)\epsilon _{3,5}+\sum _{i=6}^n \sigma _i \bigg(\bar{z}_5 \left(\epsilon _{3,5}-\epsilon _{1,3,5,i}\right)+\bar{z}_i \left(\epsilon _{1,3,5,i}-\epsilon
_{3,i}\right)\nn\\&+z_5 \left(\bar{z}_i \left(\epsilon _{5,i}-\epsilon _{1,3,5,i}\right)+\epsilon _{1,3,5,i}-\epsilon _{3,5}\right)+z_i
\left(\bar{z}_5 \left(\epsilon _{1,3,5,i}-\epsilon _{5,i}\right)+\epsilon _{3,i}-\epsilon _{1,3,5,i}\right)\bigg)
\\
\varphi\b_3=&\left(z_4-\bar{z}_4\right)
\epsilon _{3,4}+\sum _{i=6}^n \sigma _i \bigg(\bar{z}_i \left(\epsilon _{3,i}-\epsilon _{1,3,4,i}\right)+\bar{z}_4 \left(\epsilon _{1,3,4,i}-\epsilon
_{3,4}\right)\nn\\&+z_i \left(\bar{z}_4 \left(\epsilon _{4,i}-\epsilon _{1,3,4,i}\right)-\epsilon _{3,i}+\epsilon _{1,3,4,i}\right)+z_4
\left(\bar{z}_i \left(\epsilon _{1,3,4,i}-\epsilon _{4,i}\right)-\epsilon _{1,3,4,i}+\epsilon _{3,4}\right)\bigg)
\eea
By inserting these into \equref{eq: higher point generic form}, we obtain the most generic form with the kinematic constraints applied.

As the most generic form is rather complicated, let us specialize into the situation where first two momenta are incoming and the rest are outgoing. Indeed, we can show that
\footnotesize
\begin{multline}
\label{eq: higher point specific form}
\tl\cA_{J_1\dots J_n }^{\De_1,\dots,\De_n}(0,\infty,1,z_4,\dots,z_n)\evaluated_{\substack{p_{1,2}\text{ : incoming}\\p_{3,\dots,n}\text{ : outgoing}}}=
\frac{\pi}{2}\frac{\mathcal{U}(\b_i)}{\abs{-2 \left(z_5 \left(-\bar{z}_4\right)+\bar{z}_4+z_4 \left(\bar{z}_5-1\right)-\bar{z}_5+z_5\right)}}\delta\left(i(\kappa-n)+\sum\limits_{i=1}^n\lambda_i\right)
\\\x
\prod\limits_{k=6}^n\left(\int\limits_0^1d\sigma_k\sigma_k^{i\lambda_k}\right)
\left(\frac{z_5
	\bar{z}_4-z_4 \bar{z}_5+\sum _{i=6}^n 2 \sigma _i \left(\left(\bar{z}_5-\bar{z}_4\right) z_i+\left(z_4-z_5\right) \bar{z}_i+z_5 \bar{z}_4-z_4 \bar{z}_5\right)}{2 \left(z_5 \left(\bar{z}_4-1\right)-\bar{z}_4-z_4 \left(\bar{z}_5-1\right)+\bar{z}_5\right)}\right)^{i\lambda_3}
\\\x
\left(\frac{\bar{z}_5-z_5+\sum _{i=6}^n 2 \sigma _i \left(-\bar{z}_5 z_i+\left(z_5-1\right) \bar{z}_i+\bar{z}_5+z_i-z_5\right)}{2 \left(z_5
	\left(\bar{z}_4-1\right)-\bar{z}_4-z_4 \left(\bar{z}_5-1\right)+\bar{z}_5\right)}\right)^{i\lambda_4}\left(-\frac{-\bar{z}_4+z_4+\sum _{i=6}^n 2 \sigma _i \left(z_4 \left(-\bar{z}_i\right)+\left(\bar{z}_4-1\right) z_i+\bar{z}_i-\bar{z}_4+z_4\right)}{2
	\left(z_5 \left(-\bar{z}_4\right)+\bar{z}_4+z_4 \left(\bar{z}_5-1\right)-\bar{z}_5+z_5\right)}\right)^{i\lambda_5}
\\\x 
\left(\frac{1}{2}-\frac{\sum _{i=6}^n 4 \left(z_5 \left(-\bar{z}_4\right)+\bar{z}_4+z_4 \left(\bar{z}_5-1\right)-\bar{z}_5+z_5\right) \sigma _i}{2
	\left(z_5 \left(-\bar{z}_4\right)+\bar{z}_4+z_4 \left(\bar{z}_5-1\right)-\bar{z}_5+z_5\right)}\right)^{i\lambda_1}
\\\x
\Bigg(
\frac{1}{2 \left(z_5
	\left(\bar{z}_4-1\right)-\bar{z}_4-z_4 \left(\bar{z}_5-1\right)+\bar{z}_5\right)}
\Bigg[
-\bar{z}_4-z_5 \left(\bar{z}_4
\left(\bar{z}_5-2\right)+1\right)+\bar{z}_5+z_4 \left(-z_5 \bar{z}_4+\left(\bar{z}_4+z_5-2\right) \bar{z}_5+1\right)
\\
+\sum _{i=6}^n 2 \sigma _i \bigg(\left(\left(z_4-1\right) \bar{z}_4+\left(z_5 \left(\bar{z}_4-1\right)-z_4 \bar{z}_4+1\right) \bar{z}_5\right)
z_i
+\left(z_5 \left(\bar{z}_5-1\right)+z_4 \left(\left(z_5-1\right) \bar{z}_4-z_5 \bar{z}_5+1\right)\right) \bar{z}_i
\\
-2 \bar{z}_4+2
\bar{z}_5+z_5 \left(\bar{z}_5^2-z_5 \bar{z}_5+\bar{z}_4 \left(\left(z_5-2\right) \bar{z}_5+3\right)-2\right)
+z_4 \left(\left(\bar{z}_4-3\right)
\bar{z}_5-z_5 \left(\bar{z}_4+\left(\bar{z}_5-2\right) \bar{z}_5\right)+2\right)\bigg)
\Bigg]
\Bigg)^{i\lambda_2}
\\\x 
\lim\limits_{L\rightarrow\infty} 
\Amp{}{j_1\dots j_n}{1-\sum\limits_{i=1}^3\b_i-\sum\limits_{i=6}^n\s_i,\;-L^{-2}\left(\sum\limits_{i=1}^3\epsilon_2\epsilon_{i+2} \b_i\left(1+z_{i+2}\zb_{i+2}\right)+\sum\limits_{i=6}^n\epsilon_2\epsilon_i \sigma_i\left(1+z_i\zb_i\right)\right),\;\b_1,\;\b_2,\;\b_3,\;\s_6,\;\dots,\;\s_n}{0,\;L,\;1,\; z_4,\;\dots,\; z_n}
\end{multline}
\normalsize

With the equation above, we can consider  several higher point amplitudes for this momentum configuration. Below, we'll only focus on one simple case: all-plus gluon rational amplitude:
\be 
\label{eq: n point amplitude}
\cA^{\texttt{gluon}}_{+\dots +}(p_i)=-c\frac{\sum\limits_{1\le i< j< k<l\le n}\<ij\>[jk]\<kl\>[li]}{\<12\>\<23\>\cdots\<(n-1)n\>\<n1\>}\delta^4\left(\sum\limits_{i=1}^np_i^\mu\right)
\ee
where $c$ is given in footnote~\ref{footnote for c} \cite{Bern:2005hs}. This leads to
\be 
\label{eq: n point amplitude 2}
\cA^{\texttt{gluon}}_{+\dots +}(\omega_1,\dots,\w_n;z_1,\dots, z_n)=-c(-2)^{4-n}\frac{\sum\limits_{1\le i< j< k<l\le n}\e_{i,j,k,l} z_{ij}\zb_{jk}z_{kl}\zb_{li}\w_i\w_j\w_k\w_l}{z_{12}z_{23}\cdots z_{n1}\w_1\w_2\dots \w_n}\delta^4\left(\sum\limits_{i=1}^np_i^\mu\right)
\ee
for which we can immediately write
\be 
\Amp{\texttt{gluon}}{+\dots +}{\s_1\dots\s_n}{z_1\dots z_n}=-c(-2)^{4-n}\frac{\sum\limits_{1\le i< j< k<l\le n}\e_{i,j,k,l} z_{ij}\zb_{jk}z_{kl}\zb_{li}\s_i\s_j\s_k\s_l}{z_{12}z_{23}\cdots z_{n1}\s_1\s_2\dots \s_n}
\ee 
where we also see that $\ka=n$.

 Let us focus on the first term, i.e.
\be 
\label{eq: gluon one loop all plus generic}
\Amp{\texttt{gluon}}{+\dots +}{\s_1\dots\s_n}{z_1\dots z_n}=-\e_{1,2,3,4}c(-2)^{4-n}\frac{ \zb_{23}\zb_{41}}{z_{23}z_{45}\cdots z_{n1}}\prod\limits_{i=5}^n\s_i^{-1}+\text{ other terms}
\ee 
We can immediately insert this into \equref{eq: higher point specific form} and obtain
\footnotesize
\begin{multline}
\label{eq: higher point specific form}
\left(\tl\cA^{\texttt{gluon}}\right)_{+\dots + }^{\De_1,\dots,\De_n}(0,\infty,1,z_4,\dots,z_n)\evaluated_{\substack{p_{1,2}\text{ : incoming}\\p_{3,\dots,n}\text{ : outgoing}}}=\pi c(-2)^{3-n}
\frac{\mathcal{U}(\b_i)}{\abs{-2 \left(z_5 \left(-\bar{z}_4\right)+\bar{z}_4+z_4 \left(\bar{z}_5-1\right)-\bar{z}_5+z_5\right)}}\frac{ \zb_{23}\zb_{41}}{z_{23}z_{45}\cdots z_{n1}}
\\\x
\delta\left(\sum\limits_{i=1}^n\lambda_i\right)
\prod\limits_{k=6}^n\left(\int\limits_0^1d\sigma_k\sigma_k^{i\lambda_k-1}\right)
\left(\frac{z_5
	\bar{z}_4-z_4 \bar{z}_5+\sum _{i=6}^n 2 \sigma _i \left(\left(\bar{z}_5-\bar{z}_4\right) z_i+\left(z_4-z_5\right) \bar{z}_i+z_5 \bar{z}_4-z_4 \bar{z}_5\right)}{2 \left(z_5 \left(\bar{z}_4-1\right)-\bar{z}_4-z_4 \left(\bar{z}_5-1\right)+\bar{z}_5\right)}\right)^{i\lambda_3}
\\\x
\left(\frac{\bar{z}_5-z_5+\sum _{i=6}^n 2 \sigma _i \left(-\bar{z}_5 z_i+\left(z_5-1\right) \bar{z}_i+\bar{z}_5+z_i-z_5\right)}{2 \left(z_5
	\left(\bar{z}_4-1\right)-\bar{z}_4-z_4 \left(\bar{z}_5-1\right)+\bar{z}_5\right)}\right)^{i\lambda_4}\left(-\frac{-\bar{z}_4+z_4+\sum _{i=6}^n 2 \sigma _i \left(z_4 \left(-\bar{z}_i\right)+\left(\bar{z}_4-1\right) z_i+\bar{z}_i-\bar{z}_4+z_4\right)}{2
	\left(z_5 \left(-\bar{z}_4\right)+\bar{z}_4+z_4 \left(\bar{z}_5-1\right)-\bar{z}_5+z_5\right)}\right)^{i\lambda_5-1}
\\\x 
\left(\frac{1}{2}-\frac{\sum _{i=6}^n 4 \left(z_5 \left(-\bar{z}_4\right)+\bar{z}_4+z_4 \left(\bar{z}_5-1\right)-\bar{z}_5+z_5\right) \sigma _i}{2
	\left(z_5 \left(-\bar{z}_4\right)+\bar{z}_4+z_4 \left(\bar{z}_5-1\right)-\bar{z}_5+z_5\right)}\right)^{i\lambda_1}
\\\x
\Bigg(
\frac{1}{2 \left(z_5
	\left(\bar{z}_4-1\right)-\bar{z}_4-z_4 \left(\bar{z}_5-1\right)+\bar{z}_5\right)}
\Bigg[
-\bar{z}_4-z_5 \left(\bar{z}_4
\left(\bar{z}_5-2\right)+1\right)+\bar{z}_5+z_4 \left(-z_5 \bar{z}_4+\left(\bar{z}_4+z_5-2\right) \bar{z}_5+1\right)
\\
+\sum _{i=6}^n 2 \sigma _i \bigg(\left(\left(z_4-1\right) \bar{z}_4+\left(z_5 \left(\bar{z}_4-1\right)-z_4 \bar{z}_4+1\right) \bar{z}_5\right)
z_i
+\left(z_5 \left(\bar{z}_5-1\right)+z_4 \left(\left(z_5-1\right) \bar{z}_4-z_5 \bar{z}_5+1\right)\right) \bar{z}_i
\\
-2 \bar{z}_4+2
\bar{z}_5+z_5 \left(\bar{z}_5^2-z_5 \bar{z}_5+\bar{z}_4 \left(\left(z_5-2\right) \bar{z}_5+3\right)-2\right)
+z_4 \left(\left(\bar{z}_4-3\right)
\bar{z}_5-z_5 \left(\bar{z}_4+\left(\bar{z}_5-2\right) \bar{z}_5\right)+2\right)\bigg)
\Bigg]
\Bigg)^{i\lambda_2}
\\+\text{ other terms}
\end{multline}
\normalsize
As we can see, insertion of the first term in \equref{eq: gluon one loop all plus generic} into \equref{eq: higher point specific form} simply shifted $\lambda_{k\ge5}$ by $i$ and included  an overall prefactor $f(z,\zb)$. The other terms in the final result have the same property: all but four of $\lambda_k$ are shifted by $i$ and they have relative factors $f(z,\zb)$ which can be read from \equref{eq: gluon one loop all plus generic}. Therefore, for all-plus one loop gluon amplitudes, we have a summation of $\frac{n!}{4!(n-4)!}$ terms where first term is given above and the others are \emph{almost} the same, the difference being shifts in different $\lambda_k$ and the overall factor of $z_k$'s in the first line which can be extracted from \equref{eq: gluon one loop all plus generic}.

\section{Conclusion}
\label{conclusion}

In this paper, we have provided explicit construction of loop level celestial amplitudes for gluons and gravitons. We believe examples of celestial scattering amplitudes at loop level is of deep theoretical interest.

As they are rational and without any divergences, the one loop all-plus and single-minus amplitudes for Yang-Mills and gravity are natural candidates that will help in understanding the holographic properties of scattering amplitudes beyond tree level. The simplicity and subtleties of these amplitudes made them excellent candidates to study spinning celestial amplitudes beyond tree level. We computed explicit examples of four and five point of such amplitudes, and provided the integrand for a particular  $n-$point amplitude. 

There are many interesting future directions that one can consider. The study of pure Yang-Mills and gravity theory at one loop may have interesting implications for $\mathcal{N}=4$ Yang Mills and $\mathcal{N}=8$ supergravity theories. In particular, all positive helicity amplitudes that we considered in pure Yang-Mills is related to (MHV) amplitude in $\mathcal{N} = 4$ Super Yang-Mills and similarly, all positive helicity amplitudes in Einstein gravity is related to $\mathcal{N}=8$ supergravity theories \cite{Bern:1996ja}. It would be interesting to investigate these connections with the usage of celestial amplitudes technology. On a related note, it is known that there are interesting relations between scattering amplitudes of gravity and of gauge theories (see \cite{Bern:2019prr}).  Such dualities have been checked for many cases but a fundamental origin of this relation is still lacking. From a practical point of view, such dualities' most powerful applications are expected at loop level computations and it is intriguing to study such relations using celestial technology.

Another specific goal is to generalize our one-loop amplitudes in pure Yang-Mills theory and gravity beyond the cases we have considered in this paper to provide more concrete examples of the celestial CFTs. We leave all of these exciting investigations to future work.

\section*{Acknowledgement}
We are especially thankful to Dhritiman Nandan for collaboration at early stage of this work.  CC thanks Sudip Ghosh and SK wants to thank Mukunda Ghimire for conversation. SA is supported by DOE grant no. DE-SC0020318 and Simons Foundation grant 488651 (Simons Collaboration on the Nonperturbative Bootstrap).
% END

\appendix 

\section{Technical details}
\subsection{Conventions for conformal frame}
\label{sec: conformal frame}

It is well known that a conformal correlator can be rewritten in terms of conformally invariant cross ratios; for example, one can write the four point correlator as\footnote{This follows from the homogeneity of the correlator in the embedding space, i.e. \mbox{$\<\cO_1(X_1)\cdots\cO_k(\lambda X_k)\cdots\cO_n(X_n)\>=\lambda^{-\De_k}\<\cO_1(X_1)\cdots\cO_k(X_k)\cdots\cO_n(X_n)\>$}.}
\begin{subequations}
\label{eq: 4 point correlator}
\be 
\<\cO_1(x_1)\cO_2(x_2)\cO_3(x_3)\cO_4(x_4)\>=\frac{\left(x_{12}^2\right)^{\De_{31}+\De_{42}}\left(x_{23}^2\right)^{\De_{12}-(\De_3+\De_4)/2}\left(x_{31}^2\right)^{\De_{21}+\De_{43}}}{\left(x_{14}^2\right)^{\De_4}}g(u,v)
\ee 
where $\De_i$ are the scaling dimensions of the operators $\cO_i$ and where the conformal cross ratios are given as 
\be 
u=\frac{x_{12}^2x_{34}^2}{x_{13}^2x_{24}^2}\;,\quad v=\frac{x_{14}^2x_{23}^2}{x_{13}^2x_{24}^2}
\ee 
\end{subequations}
for $x_{ij}\equiv x_i-x_j$ and $2\De_{ij}\equiv \De_i - \De_j$. Here, the function $g(u,v)$ is not fixed by the conformal symmetry.\footnote{ Other information about the theory or general assumptions does constrain $g(u,v)$; for example, the whole program of conformal bootstrap is based on determining/constraining this function using (among other ingredients) operator product expansion associativity and unitarity \cite{Poland:2018epd}.}

For higher point correlators there are multiple cross ratios; the conformal moduli space of $n$ points in $d$ dimensions is given as
\be 
\label{eq: number of cross ratios}
\#\text{ of cross ratios}=\frac{m(m-3)}{2}+d(n-m)\quad\text{ for }m=\min(n,d+2)
\ee 
which becomes
\be 
\#\text{ of cross ratios for }\tl\cA_{J_1\dots J_n }^{\De_1,\dots,\De_n}=2(n-3)
\ee 
as we are interested in celestial amplitudes of $n\ge 4$ gluons. We can intuitively understand this by the following argument: given any three points, we can first use translations to fix $z_1=0$, then special conformal transformation to take $z_2\rightarrow \infty$, then dilation to bring $z_3$ to unit circle, and finally rotation to get $z_3=1$. As this exhaust all conformal transformations, $z_{n>3}$ remains unfixed, hence we have $2(n-3)$ real degrees of freedom.\footnote{To understand where \equref{eq: number of cross ratios} comes from, we recommend the nice discussion in \cite{Kravchuk:2016qvl}.}

For higher point correlators, we can generalize \equref{eq: 4 point correlator} as
\begin{subequations}
\label{eq: n point conformal correlator}
\be 
\label{eq: n point conformal correlator form}
\<\cO_1(x_1)\cO_2(x_2)\cdots\cO_n(x_n)\>=\frac{\left(x_{12}^2\right)^{\sigma_{12}}\left(x_{23}^2\right)^{\sigma_{23}}\left(x_{31}^2\right)^{\sigma_{31}}}{\prod\limits_{k=4}^n\left(x_{1k}^2\right)^{\De_k}}g(u_4,v_4;u_5,v_5;\dots ;u_n;v_n)
\ee 
for
\be 
\sigma_{12}\equiv \frac{1}{2}\left(-\De_1-\De_2+\sum\limits_{i=3}^n\De_i\right)
\;,\; 
\sigma_{23}\equiv \frac{1}{2}\left(\De_1-\sum\limits_{i=2}^n\De_i\right)
\;,\; 
\sigma_{31}\equiv \frac{1}{2}\left(-\De_1+\De_2-\De_3+\sum\limits_{i=4}^n\De_i\right)
\ee 
where the conformal cross ratios are given as 
\be
u_k\equiv\frac{x_{1k}^2x_{23}^2}{x_{13}^2x_{2k}^2}\;,\quad 
v_k\equiv \frac{x_{12}^2x_{3k}^2}{x_{13}^2x_{2k}^2}
\ee 
\end{subequations}
We note that, for $n=4$, we get back \equref{eq: 4 point correlator} from \equref{eq: n point conformal correlator} with the identification $u_4=v$ and $v_4=u$.\footnote{The reason for this inverted notation is our choice of conformal frame: as we will see below, we put the second operator at infinity whereas fourth operator is put at infinity for standard conformal frame of four points. Our choice of $u_k,v_k$ in our conformal frame match the form of $u,v$ in standard conformal frame.}

As we mentioned above, conformal transformations allow us to fix $\{x_1,x_2,x_3\}\rightarrow\{0,\infty,1\}$. In higher dimensions, we can further constrain remaining points; in $2d$, they remain as unfixed variables.  Thus \equref{eq: n point conformal correlator form} becomes
\be 
\<\cO_1(0)\cO_2(\infty)\cO_2(1)\cO_4(\w_4)\cdots\cO_n(\w_n)\>=\prod\limits_{k=4}^n\abs{\w_{k}}^{-2\De_k}
g(u_4,v_4;u_5,v_5;\dots ;u_n;v_n)
\ee 
where we implicitly used \equref{eq: putting an operator at infinity}. Extracting the function $g$ from the equation above and inserting it back into \equref{eq: n point conformal correlator form} we obtain
\be
\label{eq: pulling back from conformal frame}
\<\cO_1(z_1)\cO_2(z_2)\cdots\cO_n(z_n)\>=\abs{z_{12}}^{2\sigma_{12}}\abs{z_{23}}^{2\sigma_{23}}\abs{z_{31}}^{2\sigma_{31}}
\prod\limits_{k=4}^n\frac{\abs{\omega_k}^{2\De_k}}{\abs{z_{1k}}^{2\De_k}}\<\cO_1(0)\cO_2(\infty)\cO_2(1)\cO_4(\w_4)\cdots\cO_n(\w_n)\>
\ee
for 
\be
\w_k\w_k^*=\frac{\abs{z_{1k}}^2\abs{z_{23}}^2}{\abs{z_{13}}^2\abs{z_{2k}}^2}\;,\quad 
(1-\w_k)(1-\w_k^*)= \frac{\abs{z_{12}}^2\abs{z_{3k}}^2}{\abs{z_{13}}^2\abs{z_{2k}}^2}
\ee 

\subsection{Generalized Cramer's rule}
\label{sec: generalized Cramer's rule}

In this section, we will review the generalized Cramer's rule as derived in \cite{generalizedCramersRule}. Let's consider a system of equations of the form
\be 
\left\{
\sum\limits_{i=1}^n a_{1,i}x_i=a_{1,n+1}\;,\quad
\sum\limits_{i=1}^n a_{2,i}x_i=a_{2,n+1}\;,\quad 
\dots \;,\quad
\sum\limits_{i=1}^n a_{m,i}x_i=a_{m,n+1}
\right\}\;,\; n\ge m
\ee 
for which we can define the order$-m$ minors of  the augmented matrix as
\be 
\label{eq: matrix minors}
M_{j_1,j_2,\dots,j_m}\equiv \det \begin{pmatrix}
a_{1,j_1} & a_{1,j_2} & \dots & a_{1,j_m}\\
a_{2,j_1} & a_{2,j_2} & \dots & a_{2,j_m}\\
\dots & \dots &\dots &\dots \\
a_{m,j_1} & a_{m,j_2} & \dots & a_{m,j_m}
\end{pmatrix}
\ee 
we can then write down $x_{1,\dots,m}$ in terms of $a_{ij}$ and $x_{m+1,\dots,n}$ as
\be 
x_i=\a_{i,n+1}+\sum\limits_{j=m+1}^n\a_{i,j}x_j\;,\quad 1\le i \le m
\ee 
for
\be 
\label{eq: minor expansion coefficient}
\a_{i,j}\equiv  \frac{M_{1,2\dots,i-1,j,i+1,\dots,m-1,m}}{M_{1,2,\dots,m-1,m}}
\ee 
We hence have
\be 
\label{eq: direc delta rearrangement}
\prod\limits_{k=1}^m\delta\left(\sum\limits_{i=1}^n a_{k,i}x_i-a_{k,n+1}\right)=\frac{\prod\limits_{i=1}^m
	\delta\left(
	x_i-\a_{i,n+1}-\sum\limits_{j=m+1}^n\a_{i,j}x_j
	\right)}{\abs{M_{1,2,\dots,m}}}
\ee 

With \equref{eq: celestial amplitude from simplex coordinates 2} in mind, we can use the equation above to write down
\begin{multline}
\label{eq: delta rearrangement}
\left(\prod\limits_{k=3}^n\int\limits_0^1d\sigma_k\right)
\delta\left(\sum\limits_{i=3}^n\epsilon_i \sigma_iz_i\right)
\delta\left(\sum\limits_{i=3}^n\epsilon_i \sigma_i\zb_i\right)
\delta\left(1+\sum\limits_{i=3}^n\left(\epsilon_1\epsilon_i-1\right) \sigma_i\right)
f(\sigma_3,\dots,\sigma_n)
\\
=\frac{\mathcal{U}(\b_i)}{\abs{M_{1,2,3}}}\left(\prod\limits_{k=6}^n\int\limits_0^1d\sigma_k\right)f(\b_1,\b_2,\b_3,\s_6,\dots,\sigma_n)
\end{multline}
for
\be 
\label{eq: beta in terms of alpha}
\b_k\equiv \left\{\begin{aligned}
\a_{k,n-1}+\sum\limits_{j=4}^{n-2}\a_{k,j}\s_j &\qquad n\ge 6\\
\a_{k,n-1} & \qquad 5\ge n \ge 3
\end{aligned}\right.
\ee 
and
\be 
\label{eq: definition of function U}
\mathcal{U}(\b_i)\equiv \left\{
\begin{aligned}
	1 &\qquad 0\le \b_i \le 1
	\\
	0 & \qquad \text{ otherwise}
\end{aligned}
\right.
\ee 
where $\a$ and $M$ are given in \equref{eq: minor expansion coefficient} and \equref{eq: matrix minors} for
\be 
\label{eq: minor coefficients}
\left\{
\begin{aligned}
	a_{1,i}=&\e_{i+2}z_{i+2},\;&a_{2,i}=&\e_{i+2}\zb_{i+2},\;&a_{3,i}=&\left(\e_1\e_{i+2}-1\right)&\quad\text{ for }1\le i\le  n-2
	\\
	a_{1,n-1}=&0,\;&a_{2,n-1}=&0,\;&a_{3,n-1}=&-1&
\end{aligned}
\right\}\;, n>4
\ee 

When $n=5$ all integrals are taken care of by the Dirac-delta functions, so  \equref{eq: delta rearrangement} becomes quite simple:
\begin{multline}
\label{eq: 5 pt amplitude integrals carried out}
\left(\prod\limits_{k=3}^5\int\limits_0^1d\sigma_k\right)
\delta\left(\sum\limits_{i=3}^5\epsilon_i \sigma_iz_i\right)
\delta\left(\sum\limits_{i=3}^5\epsilon_i \sigma_i\zb_i\right)
\delta\left(1+\sum\limits_{i=3}^5\left(\epsilon_1\epsilon_i-1\right) \sigma_i\right)
f(\sigma_3,\dots,\sigma_n)
\\
=\frac{\mathcal{U}(\b_i)}{\abs{M_{1,2,3}}}f(\b_1,\b_2,\b_3)
\end{multline}

Fo $n=4$, we cannot exhaust all delta functions hence we cannot use \equref{eq: delta rearrangement}.\footnote{The situation is similar in $n=3$ case, however it needs to be treated separately due to kinematics of massless scattering. As $p_i\.p_j=0$, which follows from $(p_i+p_j)^2=p_k^2$ and $p_i^2=0$ for $i\ne j\ne k\in\{1,2,3\}$, we get \mbox{$\<ij\>[ij]=2 p_i\.p_j=0$}. In the standard $(-,+,+,+)$ metric with real momenta, $[ij]$ and $\<ij\>$ are related to each other by complex conjugation as $[ij]\propto \zb_{ij}$ and $\<ij\>\propto z_{ij}$, therefore the only consistent solution is if $\<ij\>=[ij]=0$. To circumvent this issue, one either complexify the momenta or use the metric $(-,+,-,+)$ for which $z$ and $\zb$ are real and independent variables, allowing a nontrivial solution for $\<ij\>[ij]=0$. In this paper, we  focus on $n>3$ amplitudes and do not deal with such subtleties.} We instead have
\begin{multline}
\label{eq: delta rearrangement 2}
\left(\prod\limits_{k=3}^4\int\limits_0^1d\sigma_k\right)
\delta\left(\sum\limits_{i=3}^4\epsilon_i \sigma_iz_i\right)
\delta\left(\sum\limits_{i=3}^4\epsilon_i \sigma_i\zb_i\right)
\delta\left(1+\sum\limits_{i=3}^4\left(\epsilon_1\epsilon_i-1\right) \sigma_i\right)
f(\sigma_3,\sigma_4)
\\
=\mathcal{U}(\b_i)\delta\left(z_3\zb_4-z_4\zb_3\right)f(\b_1,\b_2)
\end{multline}
where $\a$ and $M$ are given in \equref{eq: minor expansion coefficient} and \equref{eq: matrix minors} for
\be 
\label{eq: minor coefficients 2}
\left\{
\begin{aligned}
	a_{1,i}=&\e_{i+2}z_{i+2},\;&a_{2,i}=&(\e_1\e_{i+2}-1)&\quad\text{ for }1\le i\le  2
	\\
	a_{1,3}=&0,\;&a_{2,3}=&-1&
\end{aligned}
\right\}\;, n=4
\ee 

\subsection{Details for the five point amplitudes}
\label{sec: five point appendix}

The coefficients of the prescription given in \equref{eq: five point prescription} read as
\small
\be 
\label{eq: five point amplitude prescription}
a^{(5)}_{12}=  & -\sqrt{\frac{\left(-z_5 \bar{z}_4+\bar{z}_4+\left(z_4-1\right) \bar{z}_5-z_4+z_5\right) \left(\bar{z}_4+z_5 \left(\bar{z}_4
		\left(\bar{z}_5-2\right)+1\right)-\bar{z}_5+z_4 \left(z_5 \bar{z}_4-\left(\bar{z}_4+z_5-2\right) \bar{z}_5-1\right)\right) \epsilon
		_{1,2}}{\left(\epsilon _5 \left(\epsilon _4 \left(\left(-z_5 \left(\bar{z}_4-1\right)+\bar{z}_4+\left(z_4-1\right) \bar{z}_5\right) \epsilon
		_{1,3}-z_4 \left(\bar{z}_5+\epsilon _{1,3}\right)+z_5 \bar{z}_4\right)+\epsilon _3 \left(\bar{z}_5-z_5\right)\right)+\left(z_4-\bar{z}_4\right)
		\epsilon _{3,4}\right)^2}} \\
a^{(5)}_{13}= & -\sqrt{-\frac{\left(z_5 \left(\bar{z}_4-1\right)-\bar{z}_4-z_4 \left(\bar{z}_5-1\right)+\bar{z}_5\right) \left(z_5
		\bar{z}_4-z_4 \bar{z}_5\right) \epsilon _{1,3}}{\left(\epsilon _5 \left(\epsilon _4 \left(\left(-z_5
		\left(\bar{z}_4-1\right)+\bar{z}_4+\left(z_4-1\right) \bar{z}_5\right) \epsilon _{1,3}-z_4 \left(\bar{z}_5+\epsilon _{1,3}\right)+z_5
		\bar{z}_4\right)+\epsilon _3 \left(\bar{z}_5-z_5\right)\right)+\left(z_4-\bar{z}_4\right) \epsilon _{3,4}\right)^2}} \\
a^{(5)}_{23}= & \sqrt{\frac{\left(z_5 \bar{z}_4-z_4 \bar{z}_5\right) \left(\bar{z}_4+z_5 \left(\bar{z}_4
		\left(\bar{z}_5-2\right)+1\right)-\bar{z}_5+z_4 \left(z_5 \bar{z}_4-\left(\bar{z}_4+z_5-2\right) \bar{z}_5-1\right)\right) \epsilon
		_{2,3}}{\left(\epsilon _5 \left(\epsilon _4 \left(\left(-z_5 \left(\bar{z}_4-1\right)+\bar{z}_4+\left(z_4-1\right) \bar{z}_5\right) \epsilon
		_{1,3}-z_4 \left(\bar{z}_5+\epsilon _{1,3}\right)+z_5 \bar{z}_4\right)+\epsilon _3 \left(\bar{z}_5-z_5\right)\right)+\left(z_4-\bar{z}_4\right)
		\epsilon _{3,4}\right)^2}} \\
a^{(5)}_{14}= & -\bar{z}_4 \sqrt{\frac{\left(z_5-\bar{z}_5\right) \left(z_5 \left(\bar{z}_4-1\right)-\bar{z}_4-z_4
		\left(\bar{z}_5-1\right)+\bar{z}_5\right) \epsilon _{1,4}}{\left(\epsilon _5 \left(\epsilon _4 \left(\left(-z_5
		\left(\bar{z}_4-1\right)+\bar{z}_4+\left(z_4-1\right) \bar{z}_5\right) \epsilon _{1,3}-z_4 \left(\bar{z}_5+\epsilon _{1,3}\right)+z_5
		\bar{z}_4\right)+\epsilon _3 \left(\bar{z}_5-z_5\right)\right)+\left(z_4-\bar{z}_4\right) \epsilon _{3,4}\right)^2}} \\
a^{(5)}_{24}= & \sqrt{\frac{\left(z_5-\bar{z}_5\right) \left(-\bar{z}_4-z_5 \left(\bar{z}_4 \left(\bar{z}_5-2\right)+1\right)+\bar{z}_5+z_4
		\left(-z_5 \bar{z}_4+\left(\bar{z}_4+z_5-2\right) \bar{z}_5+1\right)\right) \epsilon _{2,4}}{\left(\epsilon _5 \left(\epsilon _4
		\left(\left(-z_5 \left(\bar{z}_4-1\right)+\bar{z}_4+\left(z_4-1\right) \bar{z}_5\right) \epsilon _{1,3}-z_4 \left(\bar{z}_5+\epsilon
		_{1,3}\right)+z_5 \bar{z}_4\right)+\epsilon _3 \left(\bar{z}_5-z_5\right)\right)+\left(z_4-\bar{z}_4\right) \epsilon _{3,4}\right)^2}} \\
a^{(5)}_{34}= & -\left(\bar{z}_4-1\right) \sqrt{-\frac{\left(z_5-\bar{z}_5\right) \left(z_5 \bar{z}_4-z_4 \bar{z}_5\right) \epsilon
		_{3,4}}{\left(\epsilon _5 \left(\epsilon _4 \left(\left(-z_5 \left(\bar{z}_4-1\right)+\bar{z}_4+\left(z_4-1\right) \bar{z}_5\right) \epsilon
		_{1,3}-z_4 \left(\bar{z}_5+\epsilon _{1,3}\right)+z_5 \bar{z}_4\right)+\epsilon _3 \left(\bar{z}_5-z_5\right)\right)+\left(z_4-\bar{z}_4\right)
		\epsilon _{3,4}\right)^2}} \\
a^{(5)}_{15}= & -\bar{z}_5 \sqrt{\frac{\left(z_4-\bar{z}_4\right) \left(-z_5 \bar{z}_4+\bar{z}_4+\left(z_4-1\right) \bar{z}_5-z_4+z_5\right)
		\epsilon _{1,5}}{\left(\epsilon _5 \left(\epsilon _4 \left(\left(-z_5 \left(\bar{z}_4-1\right)+\bar{z}_4+\left(z_4-1\right) \bar{z}_5\right)
		\epsilon _{1,3}-z_4 \left(\bar{z}_5+\epsilon _{1,3}\right)+z_5 \bar{z}_4\right)+\epsilon _3
		\left(\bar{z}_5-z_5\right)\right)+\left(z_4-\bar{z}_4\right) \epsilon _{3,4}\right)^2}} \\
a^{(5)}_{25}=  & \sqrt{\frac{\left(z_4-\bar{z}_4\right) \left(\bar{z}_4+z_5 \left(\bar{z}_4 \left(\bar{z}_5-2\right)+1\right)-\bar{z}_5+z_4
		\left(z_5 \bar{z}_4-\left(\bar{z}_4+z_5-2\right) \bar{z}_5-1\right)\right) \epsilon _{2,5}}{\left(\epsilon _5 \left(\epsilon _4 \left(\left(-z_5
		\left(\bar{z}_4-1\right)+\bar{z}_4+\left(z_4-1\right) \bar{z}_5\right) \epsilon _{1,3}-z_4 \left(\bar{z}_5+\epsilon _{1,3}\right)+z_5
		\bar{z}_4\right)+\epsilon _3 \left(\bar{z}_5-z_5\right)\right)+\left(z_4-\bar{z}_4\right) \epsilon _{3,4}\right)^2}} \\
a^{(5)}_{35}= & -\left(\bar{z}_5-1\right) \sqrt{-\frac{\left(z_4-\bar{z}_4\right) \left(z_4 \bar{z}_5-z_5 \bar{z}_4\right) \epsilon
		_{3,5}}{\left(\epsilon _5 \left(\epsilon _4 \left(\left(-z_5 \left(\bar{z}_4-1\right)+\bar{z}_4+\left(z_4-1\right) \bar{z}_5\right) \epsilon
		_{1,3}-z_4 \left(\bar{z}_5+\epsilon _{1,3}\right)+z_5 \bar{z}_4\right)+\epsilon _3 \left(\bar{z}_5-z_5\right)\right)+\left(z_4-\bar{z}_4\right)
		\epsilon _{3,4}\right)^2}} \\
a^{(5)}_{45}= & \left(\bar{z}_4-\bar{z}_5\right) \sqrt{-\frac{\left(z_4-\bar{z}_4\right) \left(z_5-\bar{z}_5\right) \epsilon
		_{4,5}}{\left(\epsilon _5 \left(\epsilon _4 \left(\left(-z_5 \left(\bar{z}_4-1\right)+\bar{z}_4+\left(z_4-1\right) \bar{z}_5\right) \epsilon
		_{1,3}-z_4 \left(\bar{z}_5+\epsilon _{1,3}\right)+z_5 \bar{z}_4\right)+\epsilon _3 \left(\bar{z}_5-z_5\right)\right)+\left(z_4-\bar{z}_4\right)
		\epsilon _{3,4}\right)^2}}
\ee 
\normalsize
By using these coefficients in the prescription of \equref{eq: five point prescription} for the five point gluon amplitudes given in \equref{eq: five point amplitude}, we obtain 
\begin{subequations}
\label{eq: five point result}
\scriptsize
\begin{multline}
\lim\limits_{L\rightarrow\infty} 
\Amp{\texttt{gluon}}{+++++}{1-\sum\limits_{i=1}^3\b_i,\;-L^{-2}\sum\limits_{i=1}^3\epsilon_2\epsilon_{i+2} \b_{i}\left(1+z_{i+2}\zb_{i+2}\right),\;\b_1,\;\b_2,\;\b_3}{0,\;L,\;1,\; z_4,\;z_5}
=
\frac{\bar{z}_4 \epsilon _{1,2,3,4} \left| \frac{\left(\bar{z}_4-z_4\right) \left(\epsilon _{1,5}-1\right) \epsilon
		_{3,4}+\left(z_5-\bar{z}_5\right) \left(\epsilon _{1,4}-1\right) \epsilon _{3,5}+\left(z_4 \bar{z}_5-z_5 \bar{z}_4\right) \left(\epsilon
		_{1,3}-1\right) \epsilon _{4,5}}{z_4-\bar{z}_4}\right| }{2 z_4 z_5-2 z_5^2}
\\
+\frac{\left(\bar{z}_4-1\right) \epsilon _{2,3,4,5} \left| \frac{\left(z_5-\bar{z}_5\right) \left(\epsilon _{3,5}-\epsilon _{1,3,4,5}\right)+z_4
		\left(-\epsilon _{3,4}+\bar{z}_5 \left(\epsilon _{4,5}-\epsilon _{1,3,4,5}\right)+\epsilon _{1,3,4,5}\right)+\bar{z}_4 \left(\epsilon
		_{3,4}-\epsilon _{1,3,4,5}+z_5 \left(\epsilon _{1,3,4,5}-\epsilon _{4,5}\right)\right)}{-\bar{z}_4 z_5+z_5+\bar{z}_4+z_4
		\left(\bar{z}_5-1\right)-\bar{z}_5}\right| }{2 \left(z_4-1\right) z_5}
\\
+
\frac{\left(\bar{z}_4-1\right) \bar{z}_5 \epsilon _{1,3,4,5} \left| \frac{\left(z_5-\bar{z}_5\right) \left(\epsilon _{3,5}-\epsilon
		_{1,3,4,5}\right)+z_4 \left(-\epsilon _{3,4}+\bar{z}_5 \left(\epsilon _{4,5}-\epsilon _{1,3,4,5}\right)+\epsilon _{1,3,4,5}\right)+\bar{z}_4
		\left(\epsilon _{3,4}-\epsilon _{1,3,4,5}+z_5 \left(\epsilon _{1,3,4,5}-\epsilon _{4,5}\right)\right)}{\bar{z}_4 \left(\bar{z}_5-2\right)
		z_5+z_5+\bar{z}_4-\bar{z}_5+z_4 \left(z_5 \left(\bar{z}_4-\bar{z}_5\right)-\left(\bar{z}_4-2\right) \bar{z}_5-1\right)}\right| }{2
	\left(z_4-1\right) z_5}
\\
+\frac{\left(z_5-1\right) \bar{z}_5 \epsilon _{1,2,3,5} \left| \frac{\left(\bar{z}_4-z_4\right) \left(\epsilon _{1,5}-1\right) \epsilon
		_{3,4}+\left(z_5-\bar{z}_5\right) \left(\epsilon _{1,4}-1\right) \epsilon _{3,5}+\left(z_4 \bar{z}_5-z_5 \bar{z}_4\right) \left(\epsilon
		_{1,3}-1\right) \epsilon _{4,5}}{z_5-\bar{z}_5}\right| }{2 \left(z_4-1\right) \left(z_4-z_5\right) z_5}
	+\frac{\bar{z}_5 \epsilon _{1,2,4,5} \left| \frac{\left(\bar{z}_4-z_4\right) \left(\epsilon _{1,5}-1\right) \epsilon
			_{3,4}+\left(z_5-\bar{z}_5\right) \left(\epsilon _{1,4}-1\right) \epsilon _{3,5}+\left(z_4 \bar{z}_5-z_5 \bar{z}_4\right) \left(\epsilon
			_{1,3}-1\right) \epsilon _{4,5}}{z_5 \bar{z}_4-z_4 \bar{z}_5}\right| }{2 z_5-2 z_4 z_5}
\end{multline}
\normalsize
and
\scriptsize
\begin{multline}
\lim\limits_{L\rightarrow\infty} 
\Amp{\texttt{gluon}}{-++++}{1-\sum\limits_{i=1}^3\b_i,\;-L^{-2}\sum\limits_{i=1}^3\epsilon_2\epsilon_{i+2} \b_{i}\left(1+z_{i+2}\zb_{i+2}\right),\;\b_1,\;\b_2,\;\b_3}{0,\;L,\;1,\; z_4,\;z_5}
=
-\frac{2 \epsilon _{2,3} \left| \frac{\left(z_5 \left(\bar{z}_4-1\right)-\bar{z}_4-z_4 \left(\bar{z}_5-1\right)+\bar{z}_5\right) \left(z_5
		\bar{z}_4-z_4 \bar{z}_5\right)}{\left(z_4-\bar{z}_4\right) \left(\left(\bar{z}_5-z_5\right) \left(\epsilon _{3,5}-\epsilon
		_{1,3,4,5}\right)+\bar{z}_4 \left(-\epsilon _{3,4}+z_5 \left(\epsilon _{4,5}-\epsilon _{1,3,4,5}\right)+\epsilon _{1,3,4,5}\right)+z_4
		\left(\epsilon _{3,4}-\epsilon _{1,3,4,5}+\bar{z}_5 \left(\epsilon _{1,3,4,5}-\epsilon _{4,5}\right)\right)\right)}\right| }{z_5
	\left(z_5-z_4\right)}
\\
+\frac{2 \left| \frac{\left(z_4-\bar{z}_4\right) \left(\bar{z}_4 \left(\bar{z}_5-2\right) z_5+z_5+\bar{z}_4-\bar{z}_5+z_4 \left(z_5
		\left(\bar{z}_4-\bar{z}_5\right)-\left(\bar{z}_4-2\right) \bar{z}_5-1\right)\right)}{\left(-\bar{z}_4 z_5+z_5+\bar{z}_4+z_4
		\left(\bar{z}_5-1\right)-\bar{z}_5\right) \left(\left(z_5-\bar{z}_5\right) \left(\epsilon _{3,5}-\epsilon _{1,3,4,5}\right)+z_4 \left(-\epsilon
		_{3,4}+\bar{z}_5 \left(\epsilon _{4,5}-\epsilon _{1,3,4,5}\right)+\epsilon _{1,3,4,5}\right)+\bar{z}_4 \left(\epsilon _{3,4}-\epsilon
		_{1,3,4,5}+z_5 \left(\epsilon _{1,3,4,5}-\epsilon _{4,5}\right)\right)\right)}\right| }{\bar{z}_5}
\\
-\frac{2 z_4^3 \left(z_5-1\right) \left(\bar{z}_4-\bar{z}_5\right) \epsilon _{4,5} \left| \frac{\left(z_5-\bar{z}_5\right) \left(z_5
		\left(\bar{z}_4-1\right)-\bar{z}_4-z_4 \left(\bar{z}_5-1\right)+\bar{z}_5\right)}{\left(\bar{z}_4 \left(\bar{z}_5-2\right)
		z_5+z_5+\bar{z}_4-\bar{z}_5+z_4 \left(z_5 \left(\bar{z}_4-\bar{z}_5\right)-\left(\bar{z}_4-2\right) \bar{z}_5-1\right)\right)
		\left(\left(z_5-\bar{z}_5\right) \left(\epsilon _{3,5}-\epsilon _{1,3,4,5}\right)+z_4 \left(-\epsilon _{3,4}+\bar{z}_5 \left(\epsilon
		_{4,5}-\epsilon _{1,3,4,5}\right)+\epsilon _{1,3,4,5}\right)+\bar{z}_4 \left(\epsilon _{3,4}-\epsilon _{1,3,4,5}+z_5 \left(\epsilon
		_{1,3,4,5}-\epsilon _{4,5}\right)\right)\right)}\right| }{\left(z_4-z_5\right)^2}
\end{multline}
\normalsize
\end{subequations}
One can insert these expressions into \equref{eq: five point amplitudes} to get the full celestial amplitude.

\bibliographystyle{utphys}
\bibliography{collectiveReferenceLibrary}{}
\end{document}